\documentclass[a4paper,fleqn]{cas-sc}

\usepackage[authoryear]{natbib}
\usepackage{graphicx}
\usepackage[caption=false]{subfig}
\usepackage{float}
\usepackage{algorithm}  
\usepackage{algpseudocode}
\usepackage{color}
\usepackage{setspace}
\def\tsc#1{\csdef{#1}{\textsc{\lowercase{#1}}\xspace}}
\tsc{WGM}
\tsc{QE}
\tsc{EP}
\tsc{PMS}
\tsc{BEC}
\tsc{DE}
\begin{document}
\let\WriteBookmarks\relax
\def\floatpagepagefraction{1}
\def\textpagefraction{.001}
\shorttitle{Auto-Tuning for OpenMP Dynamic Scheduling applied to FWI}
\shortauthors{Silva et al.}

\title [mode = title]{Auto-Tuning for OpenMP Dynamic Scheduling applied to Full Waveform Inversion}

\author[1]{Felipe H. S. da Silva}[type=editor, orcid=0009-0009-4739-2957]
\credit{Methodology, Software, Verification, Formal analysis, Investigation, Data Curation, Writing - Original Draft, Visualization}

\author[2]{Jo\~{a}o B. Fernandes}[orcid=0000-0001-7948-5802]
\credit{Conceptualization, Methodology, Software, Verification, Formal analysis, Investigation, Data Curation, Writing - Review \& Editing, Visualization, Supervision}

\author[2]{Idalmis M. Sardina}[orcid=0000-0003-1812-2017]
\credit{Writing - Review \& Editing}

\author[2]{Tiago Barros}[orcid=0000-0001-9665-2238]
\credit{Writing - Review \& Editing}

\author[2]{Samuel Xavier-de-Souza}[orcid=0000-0001-8747-4580]
\credit{Resources, Writing - Review \& Editing, Project administration, Funding acquisition}

\author[1]{Italo A. S. Assis}[orcid=0000-0003-4122-3115]
\credit{Conceptualization, Verification, Resources, Writing - Review \& Editing, Visualization, Supervision, Project administration, Funding acquisition}

\address[1]{Universidade Federal Rural do Semi-\'{A}rido}
\address[2]{Universidade Federal do Rio Grande do Norte}

\begin{abstract}
%Contextualization
Full Waveform Inversion (FWI) is a widely used method in seismic data processing, capable of estimating models that represent the characteristics of the geological layers of the subsurface. Because it works with a massive amount of data, the execution of this method requires much time and computational resources, being restricted to large-scale computer systems such as supercomputers. Techniques such as FWI adapt well to parallel computing and can be parallelized in shared memory systems using the application programming interface (API) OpenMP. The management of parallel tasks can be performed through loop schedulers contained in OpenMP. The dynamic scheduler stands out for distributing predefined fixed-size chunk sizes to idle processing cores at runtime. It can better adapt to FWI, where data processing can be irregular. However, the relationship between the size of the chunk size and the runtime is unknown. Optimization techniques can employ meta-heuristics to explore the parameter search space, avoiding testing all possible solutions. Here, we propose a strategy to use the Parameter Auto-Tuning for Shared Memory Algorithms (PATSMA), with Coupled Simulated Annealing (CSA) as its optimization method, to automatically adjust the chunk size for the dynamic scheduling of wave propagation, one of the most expensive steps in FWI. Since testing each candidate chunk size in the complete FWI is unpractical, our approach consists of running a PATSMA where the objective function is the runtime of the first time iteration of the first seismic shot of the first FWI iteration. The resulting chunk size is then employed in all wave propagations involved in an FWI. We conducted tests to measure the runtime of an FWI using the proposed auto-tuning, varying the problem size and running on different computational environments, such as supercomputers and cloud computing instances. The results show that applying the proposed auto-tuning in an FWI reduces its runtime by up to 70.46\% compared to standard OpenMP schedulers.
%Conclusion
\end{abstract}


\maketitle 

\printcredits
% \doublespacing

\section{Introduction}\label{introducao}

%Contextualization
Seismic surveying is a technique that involves utilizing seismic shots to emit waves and sensors to capture data from returning waves. This data is then processed to gather insights about what is in the subsurface. Seismic modeling is a part of the surveying process that can be utilized to simulate wave propagation and subsurface properties. While it can model various parameters for increased accuracy, doing so can result in more complex mathematical calculations that require intensive computational processing and time. The result of this process is what is known as calculated data.

Full Waveform Inversion (FWI) \citep{tarantola1984inversion} is one of the most used methods to estimate a velocity model. An FWI iteratively adjusts the velocity model using an optimization strategy. It aims to reduce the residual data, i.e., the difference between the observed and calculated data. FWI computes adjustments to the velocity models from the calculated and residual data.

%State the Gap: Open Questions, Restrictions, and Limitations
FWI presents challenges, such as the need for high-quality data and the computational complexity of the process. Due to its high computational demands, the execution of this algorithm in industry is restricted to large computing environments where the use of parallel programming techniques is necessary.

OpenMP (Open Multi-Processing ) \citep{OpemMP}, a library for parallel programming in shared memory systems, is a widely used tool in the parallelization of large-scale applications such as the FWI. It allows programmers to write parallel code to distribute the work of an application between multiple execution threads. Those threads are then assigned to cores of a shared memory system. By default, in OpenMP, this load is divided equally. However, in complex problems such as FWI, load imbalance may arise because some parts of the code execute uneven tasks, making some threads complete their work more quickly than others. The faster threads will become idle while waiting for the other threads to finish. Load imbalance can result in underutilization of the available hardware and, consequently, a reduction in the program's overall performance.

%State the purpose of the paper
Here, we propose a method to balance the workload among CPUs on a computing node, avoiding resource underutilization. To do this, we employ the Parameter Auto-Tuning for Shared Memory Algorithms (PATSMA) \citep{fernandes2024patsma} to automatically adjust the chunk size of the OpenMP \textit{dynamic} scheduler. The main contribution of this work is to adapt PATSMA for FWI and evaluate its performance. We compare our proposal with standard OpenMP schedulers, showing that the proposed auto-tuning reduces the FWI execution time in different computing environments and tests for several input sizes.

The rest of this paper is organized as follows: Section \ref{section:background} presents the characteristics of our target application, FWI, and the auto-tuning method we use, PATSMA. It also discusses workload distribution through loop scheduling supported by OpenMP. Next, in Section \ref{section:at}, we detail our FWI application and the proposed auto-tuning approach. We show the proposed method's results compared with the standard OpenMP schedulers in Section \ref{section:resultados}. Section \ref{section:related_work} presents the related work, and Section \ref{section:conclusoes} concludes this work.

\section{Background}\label{section:background}

This section briefly describes our target application, the FWI (Section \ref{subsection:fwi}), shows the standard OpenMP loop schedulers we compare to our approach (Section \ref{subsection:escalonadores}), and presents the auto-tuning method we use, PATSMA (Section \ref{subsection:patsma}).

\subsection{FWI}\label{subsection:fwi}

Full waveform inversion (FWI) \citep{tarantola1984inversion} is a seismic data inversion technique that aims to obtain an accurate model of the subsurface geophysical properties. This technique minimizes an objective function that measures the difference between observed and calculated data, i.e., the residual data. The gradient of the objective function indicates the search direction for the FWI optimization problem \citep{plessix2006adjoint}. The model is updated based on that gradient. Mathematically, FWI can be described as a least squares optimization problem whose objective function is
\begin{equation}
\label{eq:fwi}
\min_{\mathbf{m}\in \mathbb{R}^{n}} \frac{1}{2}\|\mathbf{d - G(m)}\|^2,
\end{equation}
where $\|.\|$ denotes the Euclidean norm; $\mathbf{m}$ is the vector that contains all the model parameters; $\mathbf{d}$ are the observed data, and $\mathbf{G}$ is the seismic modeling operator responsible for simulate the seismic acquisition in an area represented by $\mathbf{m}$.

This process is repeated until a stopping criterion is reached. Common stopping conditions are the convergence of the optimization method (the norm of the gradient is less than a predefined threshold) or the maximum number of iterations. The main steps of FWI are:

\begin{enumerate}
    \item Read observed data;
    \item Read an initial model;
    \item Compute the solution of the wave equation (calculated data) for the initial model; \label{alg:passo3fwi}
    \item Compute the objective function (Equation \ref{eq:fwi});
    \item Compute the gradient of the objective function;
    \item Update the model based on the gradient of the objective function; \label{alg:passo6fwi}
    \item Repeat steps \ref{alg:passo3fwi} to \ref{alg:passo6fwi} until the stopping criterion is reached.
\end{enumerate}

\subsection{OpenMP schedulers} \label{subsection:escalonadores}

OpenMP (Open Multi-Processing) is an Application Programming Interface (API) for parallel programming on multicore systems with shared memory \citep{OpenMP-spec}. It provides tools for programmers to develop parallel codes in C, C++, and Fortran without significant changes to the original code, using its directives to distribute application work across multiple threads.

\textit{\#pragma omp for} is a widely used OpenMP directive. That directive distributes the iterations of a loop between the threads that execute them simultaneously in different cores, allowing the program to execute these loops faster in shared memory systems. \textit{schedule} is one of the most important clauses of the directive \textit{omp for}. A clause is an instruction to change the default behavior of a directive. The clause \textit{schedule} specifies how the loop iterations must be distributed among the threads. It provides three default schedulers: \textit{static}, \textit{dynamic}, and \textit{guided}. The operation of each default scheduler is defined below.

\begin{itemize}
    \item \textit{static}: Chunks of iterations are distributed round-robin among the threads before the loop is executed. The programmer may define the size of the chunks. If the chunk size is not defined, its value is approximately $N_{i}/N_{t}$, where $N_{i}$ is the number of iterations and $N_{t}$ is the number of threads.
    \item \textit{dynamic}: During runtime, chunks of loop iterations are distributed among the threads. When a thread finishes executing its chunk, it requests a new chunk from a global queue. If the programmer does not specify the chunk size, it is $1$.
    \item \textit{guided}: Similar to \textit{dynamic}, the chunks of iterations are distributed at runtime. However, the chunk size decreases with each new request. This strategy aims to handle load imbalance better. When the programmer specifies the chunk size, no chunk can be smaller than that except the last one. If a chunk size is not defined, it is $1$.
\end{itemize}

The \textit{static} scheduler can be advantageous because static distribution guarantees that each thread will execute approximately the same number of iterations. As there is no workload redistribution overhead during execution, that strategy may lead to better performance for homogeneous hardware and iterations workload. However, if the iterations workload or the hardware is heterogeneous, static distribution may result in resource underutilization as some threads may finish before others.

Applications with heterogeneous iterations workload or running in heterogeneous hardware may perform better using the \textit{dynamic} scheduler. Because such a scheduler distributes the chunks of iterations at runtime, no thread stays idle until all the chunks are distributed. However, the program may lose performance if the chunk size is not chosen appropriately. If the chunk is too large, it can generate idle time. On the other hand, if the chunk is too small, there may be excessive chunk requests, leading to a significant increase in runtime.

The \textit{guided} scheduler aims to reduce idleness compared to the \textit{static} scheduler and management overhead compared to the \textit{dynamic} scheduler. Because the last chunks are smaller, threads will not idle long. Also, as the first chunks are larger, threads will be busy longer, reducing the number of new chunk requests. The disadvantages of the \textit{guided} scheduler are that (1) there is an additional overhead to manage the chunk size, and (2) an auto-tuning method could only control the size of the last block.

Therefore, finding an ideal chunk size is essential to achieving the best possible performance of loop scheduling of an application using OpenMP. That way, it is critical to carefully adjust the chunk size for each application, depending on the number of loop iterations, the workload characteristics, and the computing environment.

\cite{at2020italo} evaluated the performance of the \textit{dynamic} and \textit{static} schedulers using a variety of simulations in a reverse time migration (RTM) algorithm \citep{Baysal1983}. The results showed that the \textit{dynamic} scheduler with chunk size auto-tuning consistently outperformed the standard \textit{static} and \textit{guided} schedulers, with improvements of up to $33\%$ in some cases. Therefore, we used the \textit{dynamic} scheduler in this work.

\subsection{PATSMA} \label{subsection:patsma}

The Parameter Auto-Tuning for Shared Memory Algorithms (PATSMA) is a software that aims to adjust a parameter of a shared memory application using an optimization method. The application programmer can define both the parameter and the optimization method. Although PATSMA is suitable for different optimization methods, we used Coupled Simulated Annealing (CSA) \citep{csa2010samuel}, one of PATSMA's built-in optimizers.

CSA is an extension of the Simulated Annealing (SA) \citep{sa1983kirkpatrick} algorithm that uses multiple SA instances to explore the solution space simultaneously. CSA increases the search diversity by using multiple SA processes (optimizers) in different subsets of the search space. The exchange of information between the optimizers helps to diversify the search, allowing CSA to find global solutions more efficiently. 

The CSA algorithm follows the same steps of generation and acceptance as SA. Both algorithms use temperatures to control their process. However, CSA has an automatic process that controls and diversifies the search by adjusting the acceptance temperature of each optimizer. This allows some optimizers to focus on local search while others on global search.

CSA explores the solution space widely with different starting points. It generates a new solution at each iteration, which can still be accepted even if it worsens. It is effective for global optimization, but proper parameter tuning is crucial. CSA is less sensitive than SA to parameter choice and produces better quality solutions even with suboptimal tuning.

PATSMA auto-tuning is a process that involves iterative loops where a parameter affects the runtime of a specific code section within the loop. To perform the auto-tuning, the target application needs to define the code section where PATSMA will act. However, sometimes the loop progress can negatively influence the auto-tuning as the section's behavior can change during the progress. To mitigate this issue, a copy of this section can be created before the loop where we will execute the auto-tuning.

Once the section is defined, PATSMA uses the CSA to generate probable solutions for this section's use and measures the resulting runtime. After all PATSMA iterations are complete, the final solution can be sent to the original section code performer for execution.

\section{Auto-tuning applied to FWI} \label{section:at}

Performing Full Waveform Inversion (FWI) can employ an algorithm to solve the 3D wave equation through finite differences. The algorithm involves nested loops that represent the space domain, and it is executed repeatedly during FWI. These loops are in the direct propagation, reverse propagation, and gradient calculation steps. They are often parallelized using the OpenMP API and an OpenMP loop scheduler. In Section \ref{subsection:escalonadores}, we show the need to define the chunk size parameter to improve the loop scheduler's performance. In some cases, the choice of a chunk may depend on factors such as memory size, number of cores, and input size.

The Full Waveform Inversion (FWI) is an intricate algorithm that involves a varying amount of calculations depending on the space region where the calculations are made. This behavior affects the chunk size definition, which can influence FWI's execution time. 
To minimize the time to run an FWI, we propose using PATSMA with CSA to adjust the chunk size for OpenMP dynamic scheduling of parallel loops in an FWI. We use the runtime to solve the 3D wave equation as PATSMA's cost function and the chunk size for OpenMP dynamic scheduling as its parameter.

% Our FWI implementation is based on \cite{rtm2019assis}. This version of FWI uses optimal checkpointing \citep{symes2010reverse} with the library proposed by \cite{checkpoint2000griewank}. As optimal checkpointing also computes the solution of the wave equation, we also use the adjusted chunk size to schedule checkpointing loops dynamically.

% Our FWI implementation is based on \cite{rtm2019assis}. This version of FWI employs optimal checkpointing \citep{symes2010reverse} using the library proposed by \cite{checkpoint2000griewank}. Optimal checkpointing operates by storing only a subset of the intermediate simulation states during the forward propagation stage. During the reverse stage, the states that were not stored are recomputed from the available checkpoints. This method significantly reduces memory requirements, with a logarithmic increase in computational cost relative to the total number of steps. Additionally, since optimal checkpointing also computes the solution of the wave equation, we dynamically adjust the chunk size to efficiently schedule checkpointing loops.

Our implementation of FWI is based on work from \cite{mamute2025preprint}. This version of FWI employs optimal checkpointing \citep{symes2010reverse} using the library proposed by \cite{checkpoint2000griewank}. Optimal checkpointing operates by storing only a subset of the intermediate simulation states during the forward propagation stage. During the reverse stage, the states that were not stored are recomputed from the available checkpoints. This method significantly reduces memory requirements, with a logarithmic increase in computational cost relative to the total number of steps. Additionally, since optimal checkpointing also computes the solution of the wave equation, we dynamically adjust the chunk size to efficiently schedule checkpointing loops.

Algorithm \ref{alg:fwi-at} illustrates an FWI with our proposed strategy. We propose executing PATSMA only on the first time step ($t_i == 0$) of the first FWI shot (Line \ref{l:12}). PATSMA's function $autotuning()$ (Line \ref{l:13}) optimizes the chunk size for loops of the wave propagator. The resulting chunk size is used in the forward propagation (Line \ref{l:fwiat1}), reverse propagation (Line \ref{l:fwiat2}), and in the re-computation of the forward field via checkpointing (Line \ref{l:fwiat3}). We chose those three sets of loops because they represent the most computationally expensive steps for an FWI and implement the wave equation solution, allowing the reuse of the chunk size obtained by PATSMA.

\begin{algorithm} \caption{Full Wavefield Inversion with the proposed auto-tuning}
\label{alg:fwi-at}
\begin{algorithmic}[1]
\State Begin time measurement
\State Read the initial model $m_0$
\State Read the number of FWI iterations ($N_{\text{fwi}}$)
\State Read the number of time iterations ($N_{\text{s}}$)
\State Initialize other parameters for checkpointing, FWI, and PATSMA
\For {$k$ = $0$ to $N_{\text{fwi}}$}
    \State \# Begin OpenMP parallel section
    \For {all shots}
        \State Read the observed data of the shot
        \If {it is the first shot and $k$ == $0$} \label{l:12}
            \State chunk size = $autotuning()$ \label{l:13}
        \EndIf
        \For {each time iteration $t_i = {0 \cdots N_{\text{s}}-1}$}
            \State \#OpenMP parallel loop directive using dynamic scheduler with the adjusted chunk size \label{l:fwiat1}
            \For {all points in the domain of space}
                \State Compute the direct wave field for the model $m_k$
            \EndFor
            \If {$t_i$ it's a checkpoint}
                \State Save checkpoint
            \EndIf
        \EndFor
        \For {each time iteration $t_i = {N_{\text{s}}-1} \cdots 0$}
            \State \#OpenMP parallel loop directive using dynamic scheduler with the adjusted chunk size \label{l:fwiat2}
            \For {all points in the domain of space}
                \State Calculate the adjunct wavefield for the model $m_k$
            \EndFor
            \State Re-compute direct wavefield in $t_i$ from checkpoints using dynamic schedule with the adjusted chunk size \label{l:fwiat3}
            \State \#OpenMP parallel loop using static scheduler
            \State Calculate the objective function gradient
            \EndFor
        \EndFor
        \State Calculate search direction
        \State Determine the step size
        \State Update the model $m_{k+1}$
        \State \# End OpenMP parallel section
\EndFor
\State Final time measurement
\end{algorithmic}
\end{algorithm}

The initial solution (chunk size) of each CSA optimizer is chosen randomly in the range [$50$, $N_{i}/N_{t}$]. We disregard chunk sizer inferior to $50$ due to the peak overhead generated to distribute them dynamically, as shown by \cite{Fernandes2018,Barros2018}. Chunk sizes larger than the size of the standard \textit{static} distribution chunk ($N_{i}/N_{t}$) are also not taken into account because they would make the number of chunks less than or equal to the number of threads, thus forcing it to be a static distribution.

For each iteration of the CSA, each optimizer measures only the execution time of the first time step of the forward propagation using its current chunk size. According to \cite{at2020italo}, the execution time of the first time step can represent the total execution time of the propagation. This first step is executed twice, and only the elapsed time of the second repetition is recorded to avoid cache population effects. The CSA then uses those time measurements as the cost function values and generates the next set of solutions.

\section{Results} \label{section:resultados}

This section details the tests we ran to measure the performance of the proposed strategy compared to standard OpenMP loop scheduling. Subsections \ref{subs:maquinas} and \ref{subsection:parametros_testes} present the computational environments and the values of the parameters we used in our tests, respectively. In Subsection \ref{subsection:testes}, we describe the tests we performed and present and discuss their results.

\subsection{Computing Environments}\label{subs:maquinas}

We used different computational environments for the performance tests of the proposed auto-tuning, as described below:
\begin{itemize}
    \item\textbf{OPT3} - VM.Optimized3.Flex is an Oracle virtual machine with an Intel Xeon 6354 processor with 18 cores, a base frequency of 3.0 GHz, a maximum turbo frequency of 3.6 GHz, 256 GB of RAM, and 39 MB of L3 cache.
    \item \textbf{SD} - represents a node of SDumont's supercomputer located at the National Center for Scientific Computing (LNCC). This node contains 2 Intel Xeon E5-2695v2 Ivy Bridge CPUs (12c @2.4GHz) totaling 24 cores, 64 GB of RAM, and 30 MB of L3 cache.
    \item \textbf{NPAD} - NPAD represents a node of the NPAD supercomputer located at the Federal University of Rio Grande do Norte (UFRN). This node contains 2 Intel Xeon CPU E5-2683 v4 @ 2.10GHz CPUs, for a total of 32 cores, 512 GB of RAM, and 40 MB of L3 cache.
    \item \textbf{STD3} - VM.Standard3.Flex is an Oracle virtual machine with an Intel Xeon Platinum 8358 processor with 32 cores, a base frequency of 2.6 GHz, a maximum turbo frequency of 3.4 GHz, 512 GB of RAM, and 48 MB of L3 cache.
    \item \textbf{STDE4} - VM.Standard.E4.Flex is an Oracle Cloud virtual machine with an AMD EPYC 7J13 processor with 64 cores, a base frequency of 2.55 GHz, a maximum boosting frequency of 3.5 GHz, 1024 GB of RAM, and 256 MB of L3 cache.
    \item \textbf{DENSE} - BM.DenseIO.E4.128 is an Oracle Cloud virtual machine with an AMD EPYC 7J13 processor with 128 cores, a base frequency of 2.55 GHz, a maximum boosting frequency of 3.5 GHz, 2048 GB of RAM, and 256 MB of L3 cache.
\end{itemize}

\subsection{Test parameters} \label{subsection:parametros_testes}
In terms of the FWI parameterization, for all tests conducted here, we used a peak frequency ($f_{peak}$) of $10 Hz$, a time sampling interval of $1 ms$, a total of $2458$ time steps, a spatial resolution of $\Delta x1 = \Delta x2 = \Delta x3 = 10 m$, and an absorbing boundary thickness of $25$ grid points in all directions of the 3D mesh.

\begin{figure}
    \centering
    \includegraphics[width=0.7\linewidth]{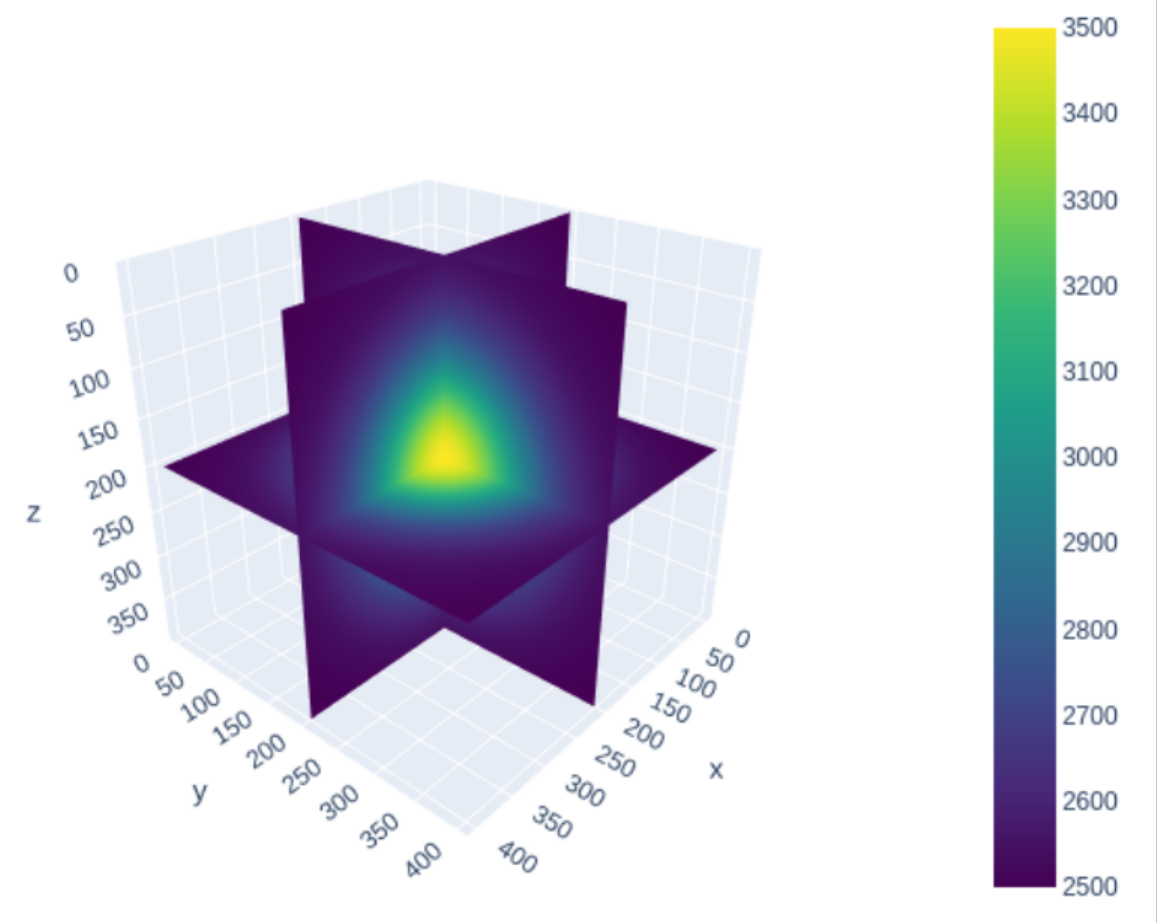}
    \caption{Velocity model with dimensions $n1 = n2 = n3 = 400$.}
    \label{fig:model-velocity}
\end{figure}

We used three true velocity models with dimensions $(n_1, n_2, n_3) = (100, 400, 400)$, $(200, 400, 400)$, and $(400, 400, 400)$, where $n_1$, $n_2$, and $n_3$ represent the number of samples along the spatial axes $x_1$, $x_2$, and $x_3$, respectively. The model with $n_1 = 400$, shown in Fig.~\ref{fig:model-velocity}, served as the base model. It was constructed using a spherical Gaussian perturbation embedded in a homogeneous background. The background velocity is $2500,\text{m/s}$, and the perturbation reaches a maximum velocity of $3500,\text{m/s}$ at the sphere's core. The two smaller models were generated by extracting subvolumes from this full model, taking only the first $n_1$ slices along the $x_1$ axis.

These true models were then used to generate synthetic seismic data, which served as observed input for the Full Waveform Inversion (FWI). The initial models used in the inversion were homogeneous with a constant velocity of $2500,\text{m/s}$, lacking any internal perturbations.

Table \ref{tab:csa-param} presents the parameters used in our experiments. The selection of the number of CSA optimizers, $m$, is extensively analyzed in \cite{gonccalves2018parallel}, where it is shown that the optimal values typically range between $m = 4$ and $m = 10$ to minimize various functions. In this work, we set $m = 4$ to reduce the computational overhead of PATSMA while maintaining the CSA performance. For the initial acceptance temperature, we adopted one of the values suggested in \cite{csa2010samuel}, $T_0^{ac}=0.9$. Furthermore, \cite{at2020italo} reports experiments that explored different values for the number of iterations, $N$, and the initial generation, ${T_0^{ac}}$. Based on their findings, we selected $N = 40$ and ${T_0^{ac}} = 100$, as these parameters yielded the shortest execution times in all test cases.

\begin{table}
\centering
\begingroup
\setlength{\tabcolsep}{10pt} % Default value: 6pt
\renewcommand{\arraystretch}{1.5} % Default value: 1
\caption{CSA parameters where ${T_0^{gen}}$ and ${T_0^{ac}}$ are the initial generation and acceptance temperatures, respectively, $N$ is the total number of iterations, and $m$ is the number of optimizers.}
\label{tab:csa-param}
\begin{tabular}{|l|l|l|l|}
\hline

 ${T_0^{gen}}$  & ${T_0^{ac}}$  & $N$   & $m$  \\ \hline
 $100$          & $0.9$         & $40$  & $4$  \\ \hline

\end{tabular}
\endgroup
\end{table}

\subsection{Performance analysis} \label{subsection:testes}

The first set of tests aims to verify whether it is possible to reuse the chunk size adjusted for the first shot for the next ones. The chosen model size was $(n1, n2, n3) = (200, 400, 400)$. Initially, we measured the time per shot for executions with $16$ shots. The tests were conducted on the NPAD machine.

\begin{figure}
    \centering
    \includegraphics[width=.7\linewidth]{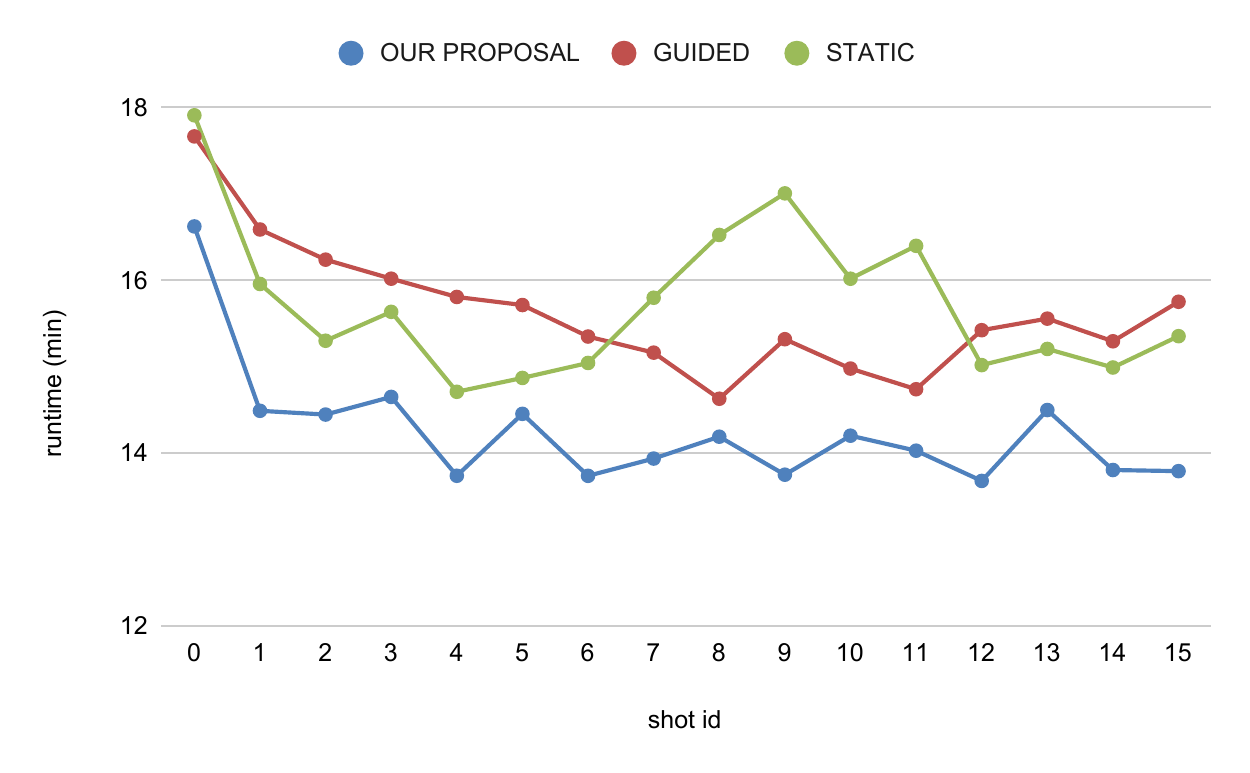} 
    \caption{Runtime per shot for FWI using the proposed auto-tuning compared to the OpenMP \textit{static} and \textit{guided} schedulers on the NPAD machine, with $16$ shots. For the OpenMP schedulers, the chunk size was not explicitly set. The velocity model size for these tests was $(n1,n2,n3)=(200,400,400)$.}
    \label{fig:time-per-shot}
\end{figure}

% Fig. \ref{fig:time-per-shot} shows that our proposal exhibits better execution time compared to the standard Static and Guided schedulers, even in the first shot, which has an overhead due to the calculation of the chunk size. Although the overhead in the first shot is present, the performance gain stabilizes after this initial execution.
Fig. \ref{fig:time-per-shot} shows that our proposal achieves better execution times compared to the standard Static and Guided schedulers. The first shot presents an initial overhead due to the computation of the chunk size; however, the performance stabilizes in subsequent executions. The static scheduler exhibits greater variability in execution time across different shots, as it is more sensitive to changes in the computational distribution. In contrast, the Guided scheduler also experiences variations but benefits from a degree of adaptation, mitigating extreme fluctuations. On the other hand, the dynamic scheduler with PATSMA demonstrates higher adaptability, maintaining more stable execution times despite changes in shot distribution.

In another set of experiments aimed at verifying the gain in full execution time per shot, we examined whether the chunk size adjusted for the first shot could be reused in subsequent executions. The chosen model size was $(n1, n2, n3) = (200, 400, 400)$, and we used $1$, $2$, $4$, $8$, $16$, $32$, $64$, and $128$ shots. These tests were performed on the NPAD machine, with each value representing a median of $5$ points.

\begin{figure}
    \centering
    \includegraphics[width=.7\linewidth]{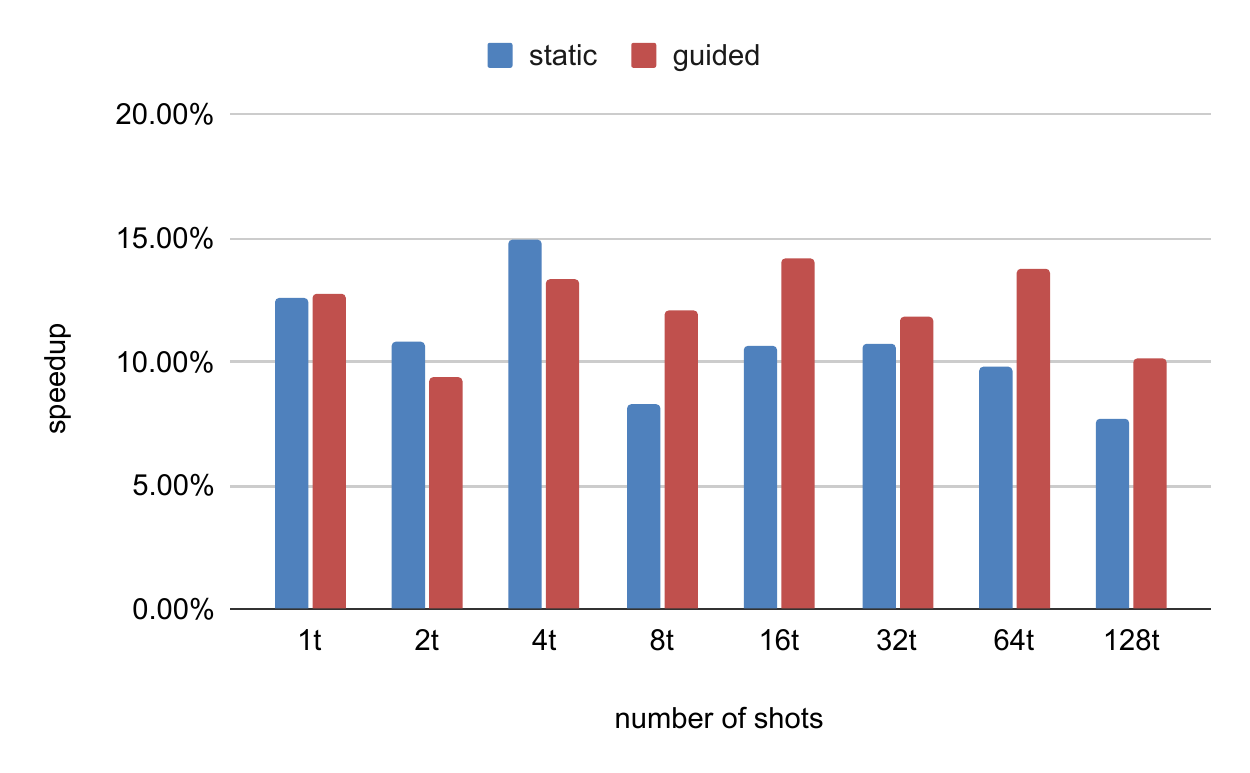} 
    \caption{Speedup for FWI using the proposed auto-tuning compared to the OpenMP \textit{static} and \textit{guided} schedulers on the NPAD machine, with $1$, $2$, $4$, $8$, $16$, $32$, $64$ and $128$ shots. For the OpenMP schedulers, the chunk size was not explicitly set. The velocity model size for these tests was $(n1,n2,n3)=(200,400,400)$. Each point is a median of five executions.}
    \label{fig:varia_tiro}
\end{figure}

Fig. \ref{fig:varia_tiro} shows that in all cases, the proposed strategy demonstrated better performance, with speedups of up to $14.1\%$ and $14.96\%$ when compared to the standard OpenMP static and guided schedulers, respectively. As we increased the number of shots, we observed that the performance of the auto-tuning strategy remained stable, as its overhead did not increase and remained similar to that of the first shot. In this set of tests, the overhead of the proposed method remained below $1.2\%$ in all cases. We conducted tests with different numbers of shots and observed that PATSMA consistently outperformed the Static and Guided schedulers. As shown in Fig. \ref{fig:time-per-shot}, the execution time for the static scheduler varied significantly depending on the shot, probably due to the influence of the shot positions in the 3D model grid, which affects computational distribution. The Guided scheduler exhibited some level of adaptation, reducing extreme fluctuations but still showing variations. In contrast, PATSMA maintained more stable performance in all cases tested, strengthening its effectiveness in handling variations in shot distribution. 

% We conducted tests for different numbers of shots, verifying that the performance gain for PATSMA remained constant. The reason for the gain not stabilizing in static might be related to the position of the shots in the 3D model grid, which could influence performance. Despite this, PATSMA showed better performance in all cases.

\begin{figure}
    \centering
    \subfloat[]{\includegraphics[width=.7\linewidth]{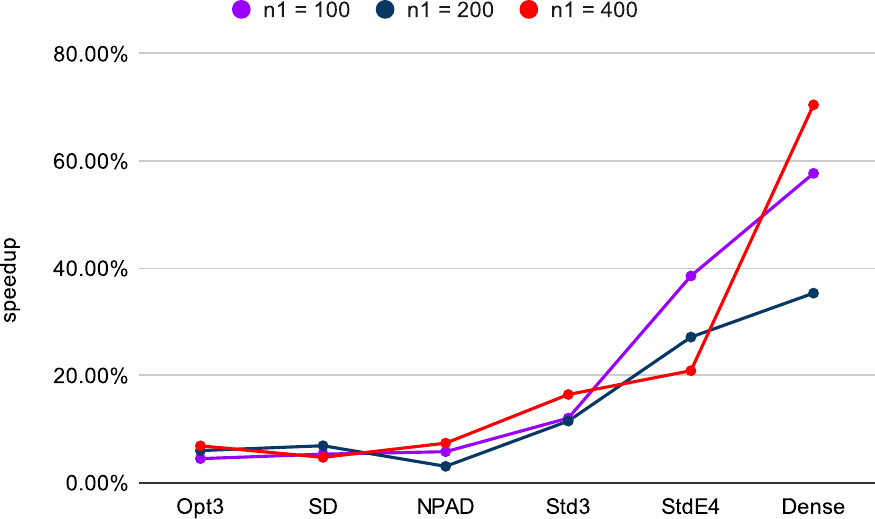}\label{subfig:speedup-machines-static}}
    
    \subfloat[]{\includegraphics[width=.7\linewidth]{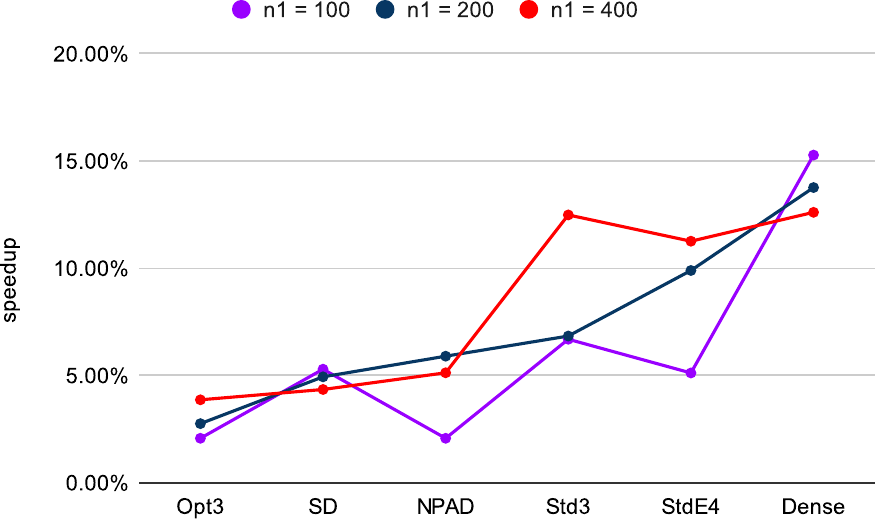}\label{subfig:speedup-machines-guided}} 
    \caption{Single-shot FWI speedup using the proposed auto-tuning compared to OpenMP (a) \textit{static} and (b) \textit{guided} schedulers on $5$ machines (OPT3, SD, NPAD, STD3, STDE4, and DENSE), for three input sizes, $(n1,n2,n3)=(100,400,400)$, $(200,400,400)$, and $(400,400,400)$. For OpenMP schedulers, the chunk size was not explicitly set. Each point is a median of at least five runs.}
    \label{fig:speedup-machines}
\end{figure}

In another set of experiments, we measured the performance of the proposed strategy as we increased the problem size and computational resources (number of cores and amount of L3 cache). Fig. \ref{fig:speedup-machines} illustrates the performance of the proposed auto-tuning compared to the standard OpenMP schedulers \textit{static} and \textit{guided} on the set of six computational environments mentioned in Subsection \ref{subs:maquinas}, varying $n1=\{100,200,400\}$. For the standard OpenMP schedulers, the chunk size was not explicitly set. Each point in the graph is a median of five samples. Fig. \ref{fig:speedup-machines} shows that the proposed method had a superior performance in all scenarios. It also shows that, in most cases, the more cores a machine has, the higher the obtained speedup. The reason is that the higher the number of cores, the smaller the search space, implying fewer possible solutions (see Section \ref{section:at}). That way, CSA is more likely to find a near-optimal solution.

As we can see, the speedups related to the \textit{static} scheduler (Fig. \ref{subfig:speedup-machines-static}) are higher than those related to the \textit{guided} scheduler (Fig. \ref{subfig:speedup-machines-guided}). A poor performance of the \textit{static} scheduler for large problem sizes can explain that phenomenon. The larger the problem size for the default \textit{static} scheduler, the larger its chunk size (see Section \ref{subsection:escalonadores}). That means reduced data locality and, consequently, reduced cache reuse. Compared to the \textit{static} scheduler, the FWI using the proposed strategy obtained a speedup of up to $70.46\%$.

Also, the wavefield is initially zero for each of the FWI's wave propagations. As the wave propagation simulation runs, only a subset of the domain has non-zero wave pressure values. Because we use compilation optimization flags, the processor might run operations with zero values faster.
When using the \textit{static} scheduler, each thread is assigned a fixed block of iterations to process, regardless of the time it takes to execute each iteration. Since blocks' load may vary, each block's processing time may differ. If a thread is faster than the others, it can become idle while waiting for the other threads to finish their work. Therefore, the performance of the \textit{static} scheduler is dominated by the slowest thread.

On the other hand, \textit{guided} scheduler reduces threads' idle time. Chunks are distributed dynamically, and by the end of the execution, chunk sizes are smaller. That way, the maximum idle time a thread may have is the time for processing the last chunk. \textit{guided} scheduler can be more efficient than \textit{dynamic} scheduler in cases where the ideal chunk size is unknown. However, the wave propagation behavior may affect its performance. Should the first chunk (the largest) contain most of the non-zero wavefield, it will dictate the performance. In addition, \textit{guided} scheduler has an overhead in managing the chunk size, which does not occur with \textit{dynamic} scheduler. FWI using the proposed approach had a speedup of up to $15.27$\% compared to using \textit{guided}.

In our last set of experiments, we compared the runtime of an FWI using the proposed auto-tuning, the OpenMP \textit{static} and \textit{guided} scheduler with their default chunk size values, and the OpenMP \textit{dynamic}, \textit{static} and \textit{guided} schedulers with the chunk size defined by \cite{auto4omp2022}:

\begin{equation}
chunk = \left\lfloor \frac{N}{2^{f} \times 2 P}\right\rfloor,\end{equation}
where
\begin{equation}
f = \left\lfloor log_{2} \left(\frac{N}{P}\right) \times {\frac{1}{\phi }} \right\rfloor,
\end{equation}
$N$ denotes the number of loop iterations, $P$ is the number of OpenMP threads, and $\phi = 1.618$. We ran this set of experiments in SD and NPAD machines.

\begin{figure}
    \centering
    \subfloat[]{\includegraphics[width=.6\linewidth]{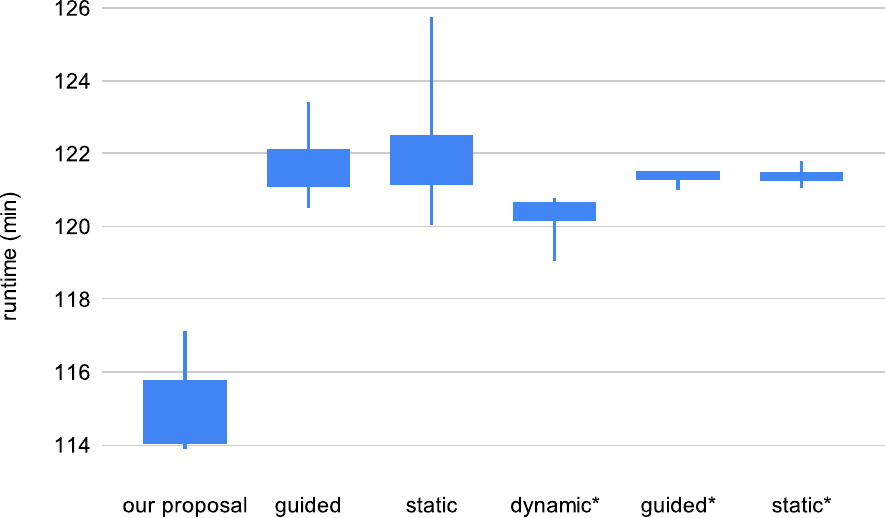}\label{subfig:mohammed22-n1100-sdumont}}
    
    \subfloat[]{\includegraphics[width=.6\linewidth]{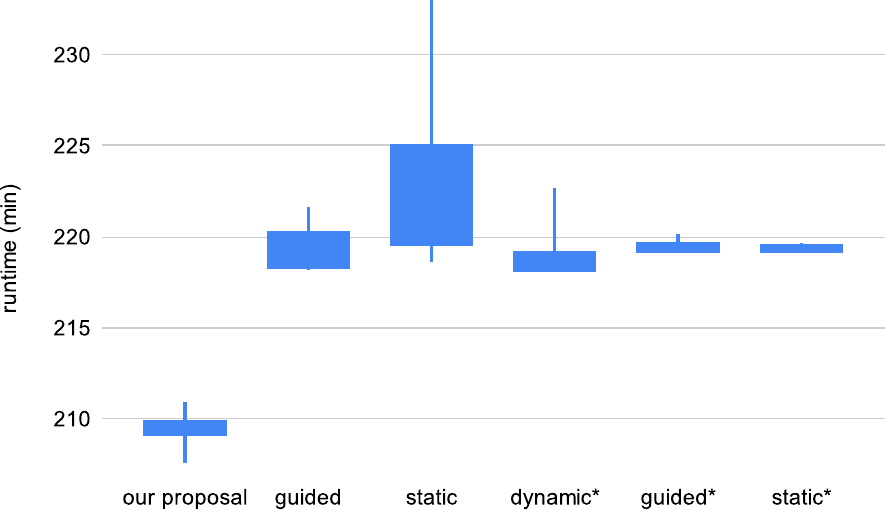}\label{subfig:mohammed22-n1200-sdumont}}
    
    \subfloat[]{\includegraphics[width=.6\linewidth]{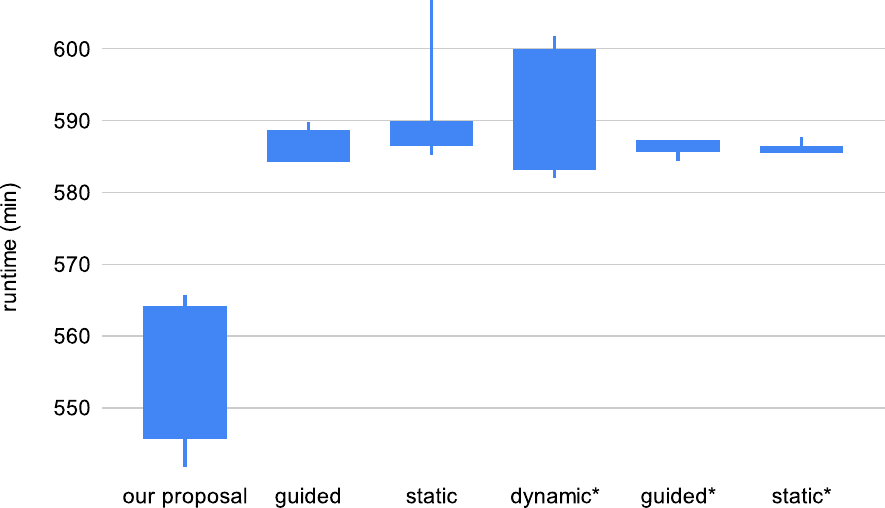}\label{subfig:mohammed22-n1400-sdumont}}
    \caption{Single-shot FWI runtime using the proposed auto-tuning compared to the default OpenMP \textit{static} and \textit{guided} schedulers, and the OpenMP \textit{dynamic}, \textit{static}, and \textit{guided} using the chunk size proposed by \cite{auto4omp2022} (marked with *). This set of experiments was performed on the SD machine for three input sizes, $(n1,n2,n3) = $ (a) $(100,400,400)$, (b) $(200,400,400)$, and (c) $(400,400,400)$. Each point is a median of five runs.}
    \label{fig:mohammed22-sdumont}
\end{figure}

Fig. \ref{fig:mohammed22-sdumont} shows the FWI runtimes measured in the SD machine. For all input sizes, the median runtime of the FWI using the OpenMP \textit{dynamic} scheduler with the chunk size defined by \cite{auto4omp2022} was lower than using the OpenMP default \textit{static} and \textit{guided} schedulers. Also, the FWI using our proposed method presented the lowest runtimes in all scenarios tested in this set of experiments being $3.75\%$, $3.88\%$, and $3.86\%$ faster than the FWI using the OpenMP \textit{dynamic} scheduler with the chunk size defined by \cite{auto4omp2022} for $n1=100$, $n1=200$, and $n1=400$, respectively.

\begin{figure}
    \centering
    \subfloat[]{\includegraphics[width=.6\linewidth]{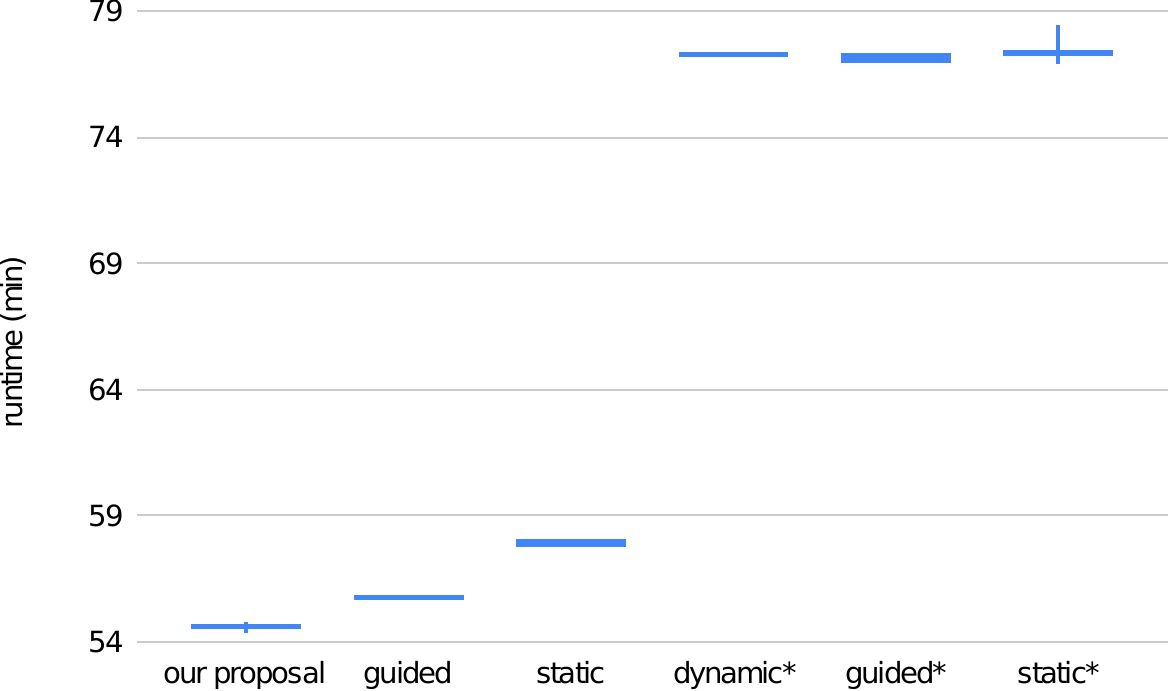}\label{subfig:mohammed22-n1100-npad}}
    
    \subfloat[]{\includegraphics[width=.6\linewidth]{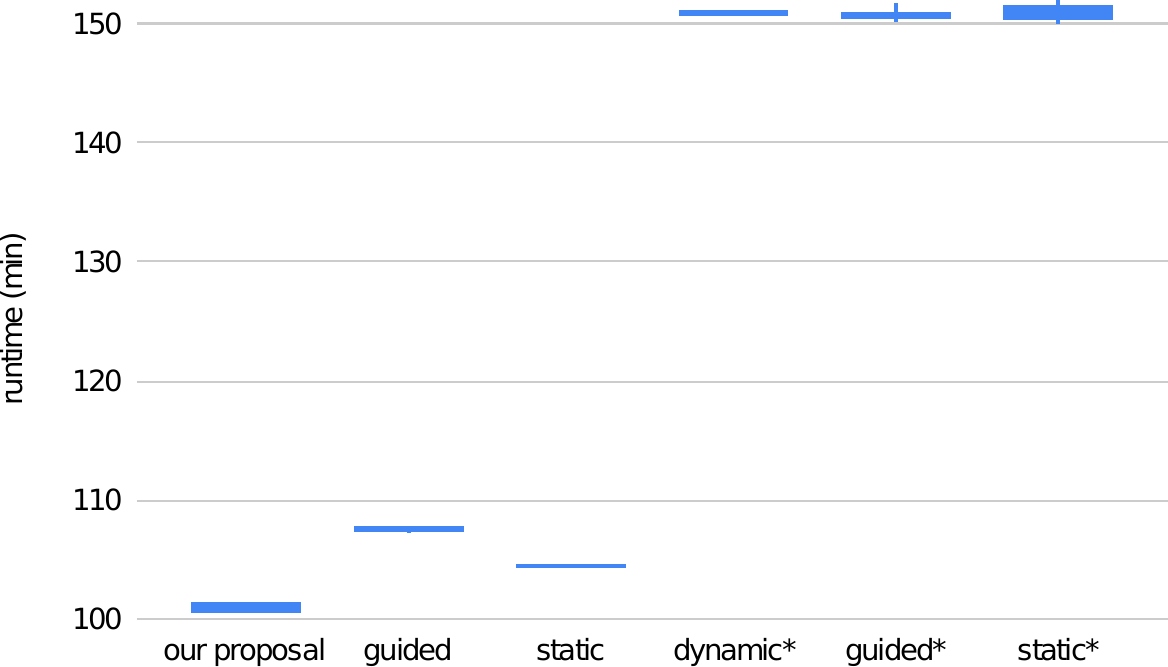}\label{subfig:mohammed22-n1200-npad}}
    
    \subfloat[]{\includegraphics[width=.6\linewidth]{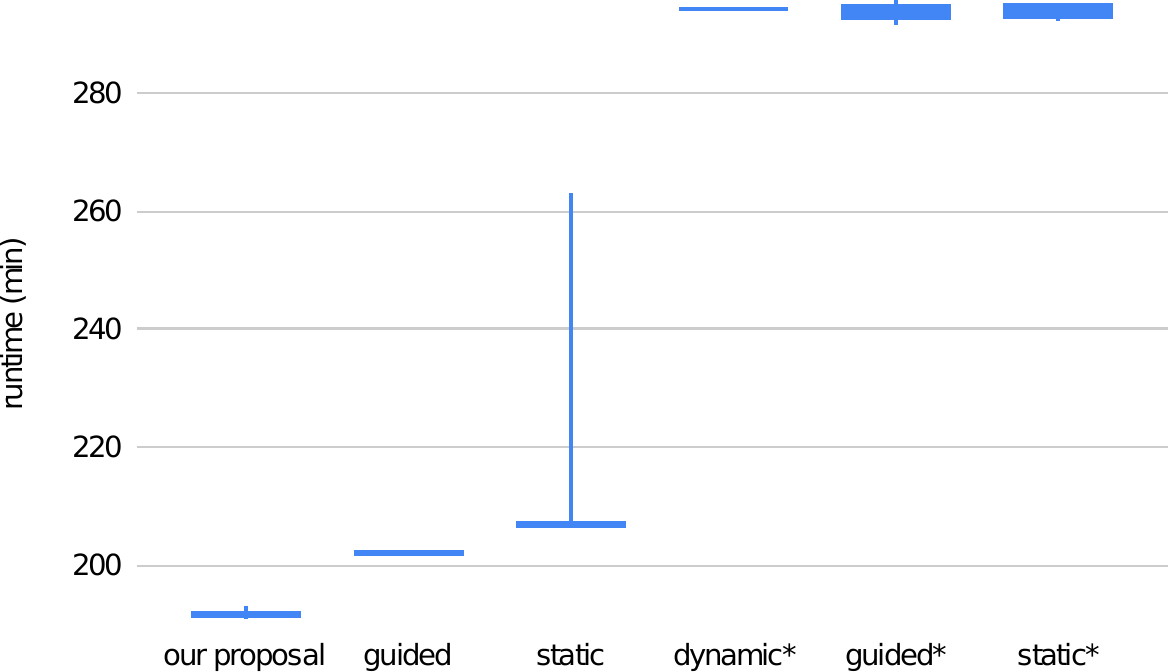}\label{subfig:mohammed22-n1400-npad}}
    \caption{Single-shot FWI runtime using the proposed auto-tuning compared to the default OpenMP \textit{static} and \textit{guided} schedulers, and the OpenMP \textit{dynamic}, \textit{static}, and \textit{guided} using the chunk size proposed by \cite{auto4omp2022} (marked with *). This set of experiments was performed on the SD machine for three input sizes, $(n1,n2,n3) = $ (a) $(100,400,400)$, (b) $(200,400,400)$, and (c) $(400,400,400)$. Each point is a median of five runs.}
    \label{fig:mohammed22-npad}
\end{figure}

Fig. \ref{fig:mohammed22-npad} shows the FWI runtimes measured in the NPAD machine. For all input sizes, the median runtime of the FWI with the chunk size defined by \cite{auto4omp2022} was higher than using the OpenMP default \textit{static} and \textit{guided} schedulers. Also, the FWI using our proposed method presented the lowest runtimes in all scenarios tested in this set of experiments being $2.07\%$, $3.05\%$, and $5.12\%$ faster than the fastest FWI using the default OpenMP \textit{dynamic} or \textit{static} schedulers for $n1=100$, $n1=200$, and $n1=400$, respectively.

\section{Related Work} \label{section:related_work}

% NÃO É PARA APLICAÇÃO
Auto-tuning has been widely studied. Some authors present methods to adjust workflows automatically. \cite{at2020hose} proposed an auto-tuning method to improve the data flow of stream processing applications based on continuously collected runtime information (e.g., available resources and incoming traffic rate). \cite{Shu2021} proposed the Component-based Ensemble Active Learning (CEAL). This auto-tuning method combines machine learning techniques with knowledge of the in-situ workflow structure to enable automated workflow configuration with limited performance measurements. Our proposal aims to autotune applications, unlike \cite{at2020hose,Shu2021}.

% É PARA APLICAÇÃO, MAS NÃO É PARA OTIMIZAÇÃO DE CÓDIGO
Application autotune has been exploited from several perspectives. \cite{Padoin2014,Padoin2017} employ load balancing techniques to control processor frequency, improving the energy efficiency of parallel applications on multicore systems. Both works use processor frequency tuning techniques to slow down less loaded cores, saving energy. \cite{at2020Hiago} presents AtTune, a heuristic-based framework to optimize the parallel execution of applications, tuning the number of processes/threads and CPU frequency levels. \cite{at2021bez} proposes to use machine learning techniques to automatically detect I/O access patterns at runtime for applications in High-Performance Computing (HPC) platforms. Based on those patterns, it is possible to dynamically adjust I/O parameters and allocate resources more efficiently to improve the system's overall performance. \cite{at2021bagbaba} presented an auto-tuning approach based on I/O monitoring and predictive modeling. This approach can find a set of I/O parameter values in a given system and application use case. This auto-tuning is applied to parallel I/O stacks in distributed systems using the MPI-IO ROMIO library. \cite{shaman2021robert} developed an optimization tool called SHAMAN, which provides a ready-to-use web application to perform black-box auto-tuning of custom computer components in a distributed system for a specific application executed through the Slurm workload manager. Different from \cite{Padoin2014,Padoin2017,at2020Hiago,at2021bez,at2021bagbaba,shaman2021robert}, our auto-tuning works on code-level optimization.

% É PARA OTIMIZAÇÃO DE CÓDIGO, MAS NÃO PARA QUALQUER ARQUITETURA DE MEMÓRIA COMPARTILHADA
Several works proposed approaches for code-level optimization auto-tuning. \cite{Kale2014} proposed to improve scheduling by combining static and dynamic approaches in symmetric multiprocessing (SMP) machines, emphasizing spatial locality. Different from \cite{Kale2014}, we propose an auto-tuning for any parallel application using OpenMP.

% É PARA OTIMIZAÇÃO DE CÓDIGO EM QUALQUER ARQUITETURA DE MEMÓRIA COMPARTILHADA, MAS PRECISA RODAR A APLICAÇÃO INTEIRA
\cite{Katagiri2014,Katagiri2015} presented ppOpen-AT, a framework for directive-driven code optimization such as loop split and loop fusion. \cite{hiperbot2020menon} presented HiPerBOt, a configuration selection framework based on Bayesian optimization, to identify application and platform-level parameters that result in high-performance configurations. \cite{Gptune2021liu} present GPTune, an auto-tuning based on multitask learning and Bayesian optimization suitable for tuning exascale application codes. \cite{kimovski2021autotuning} developed a machine learning-based framework to optimize exascale applications. The AutoTuner can adjust the application's parameters to improve performance when detecting an anomaly. \cite{at2021roy} presented Bliss, an algorithm that uses Bayesian optimization to find the near-optimal parameter configuration. Bliss must run the application a few times to determine the cost function. \cite{at2020milani} proposed policies to choose the best among user-defined code versions automatically. In common, \cite{Katagiri2014,Katagiri2015,hiperbot2020menon,Gptune2021liu,kimovski2021autotuning,at2021roy,at2020milani} need to run the application under a variety of different configurations to learn the correlation between an application's parameters and its performance, which can be prohibitive for computationally expensive applications such as FWI. We propose using the time to compute the first iteration of the laplacian as an estimative of the time of the FWI.

% É PARA OTIMIZAÇÃO DE CÓDIGO EM QUALQUER ARQUITETURA DE MEMÓRIA COMPARTILHADA, MAS NÃO É PARA ESCALONAMENTO DE LAÇOS
\cite{at2020sakurai} proposed two auto-tuning functions for heterogeneous computing architectures. The first function may perform loop transformations to adjust the position of directives in OpenMP loops automatically. The second function dynamically changes the number of threads. \cite{at2018bak} introduced a load-balancing method based on integrating Charm++ and OpenMP to distribute user-created tasks dynamically. \cite{at2020kruse} addressed compiler performance optimization through a loop transformation auto-tuning. Their proposal is based on search trees to explore the loop transformation space and uses pragmas inserted in the source code to apply these transformations. Our proposal aims to optimize parallel loops scheduling, unlike \cite{at2020sakurai,at2018bak,at2020kruse}.

% É PARA OTIMIZAÇÃO DE CÓDIGO EM QUALQUER ARQUITETURA DE MEMÓRIA COMPARTILHADA, É PARA ESCALONAMENTO DE LAÇOS, MAS É ESTÁTICO
Many authors proposed auto-tuning methods suitable for optimizing parallel loops scheduling. \cite{at2022dutta} presented preliminary work on using deep machine learning to identify loop patterns and adjust loop parameters with OpenMP. As \cite{at2022dutta} use machine learning, it is necessary to train it before execution. \cite{at2021wood} introduced Artemis, an online framework that dynamically tunes the execution of parallel regions by training optimizing models. Our proposed method requires no training, unlike \cite{at2022dutta,at2021wood}.

% É PARA OTIMIZAÇÃO DE CÓDIGO EM QUALQUER ARQUITETURA DE MEMÓRIA COMPARTILHADA, É PARA ESCALONAMENTO DE LAÇOS, MAS É ESTÁTICO
\cite{at2020seiferth} introduced an approach that autotunes applications by classifying optimization approaches based on a performance analytical model. \cite{Andreolli2014,Andreolli2015} brought an approach to automatically tune seismic applications by compiling and executing each set of parameters chosen by a genetic algorithm, including load chunk size and compilation flags. \cite{at2020seiferth,Andreolli2014,Andreolli2015} methods perform auto-tuning statically. That way, they may encounter a system state different from the runtime, as aspects such as memory availability, resource usage and access, workload, and cores' load may vary over time. On the other hand, our method performs auto-tuning dynamically at runtime, allowing it to adjust the application for the system's current state.

% É PARA OTIMIZAÇÃO DE CÓDIGO EM QUALQUER ARQUITETURA DE MEMÓRIA COMPARTILHADA, É PARA ESCALONAMENTO DE LAÇOS, MAS USA WORK STEALING (OVERHEAD MAIOR PARA GERENCIAR FILA)
\cite{ich2022booth} presented iCh, an auto-tuning method for OpenMP loop schedulers (guided and dynamic) that uses low-cost heuristics to reduce load imbalance. iCh uses on-demand work-stealing as its primary load-balancing mechanism and a distribution system with local queues for threads. By using work-stealing, iCh may have extra parallel overhead as it has to manage queues of tasks and thread synchronization. On the contrary, our proposal uses the OpenMP dynamic schedule.

% É PARA OTIMIZAÇÃO DE CÓDIGO EM QUALQUER ARQUITETURA DE MEMÓRIA COMPARTILHADA, É PARA ESCALONAMENTO DE LAÇOS, MAS TEM CUSTO EXTRA PARA GERENCIAR ESPAÇO DE BUSCA
\cite{at2021rash} presented ATF (Automatic Tuning Framework), a general approach for automatically tuning programs with interdependent tuning parameters. It focuses on optimizing the efficiency of automatic tuning by providing optimized mechanisms for generating, storing, and exploring the search space of parameters. ATF observes past constraints to generate a search space, while in our proposal, the search field is defined in advance without the need for the generation step. ATF uses a multidimensional search strategy, while we use CSA, which, as a metaheuristic, seeks to reduce the search space.

% É PARA OTIMIZAÇÃO DE CÓDIGO EM QUALQUER ARQUITETURA DE MEMÓRIA COMPARTILHADA, É PARA ESCALONAMENTO DE LAÇOS, MAS CÁLCULO DO CHUNK SIZE IGNORA DIVERSOS FATORES
\cite{auto4omp2022} introduced three methods for automated scheduler selection and an automatic chunk size adjustment for the OpenMP \textit{auto} scheduler. They provide better performance by adapting to unpredictable variations in the application and system during execution. \cite{auto4omp2022} propose an equation to compute the chunk size based on the number of threads and the problem size. Doing so may disregard other essential factors, such as hardware architecture and software behavior (e.g., memory access pattern). Using PATSMA with CSA to optimize the chunk size, our proposal will more likely consider all relevant factors. In fact, for the experiments shown in Fig. \ref{fig:mohammed22-sdumont}, an FWI performed better using a chunk size obtained with our method rather than the chunk size proposed by \cite{auto4omp2022}.

% É PARA OTIMIZAÇÃO DE CÓDIGO EM QUALQUER ARQUITETURA DE MEMÓRIA COMPARTILHADA, É PARA ESCALONAMENTO DE LAÇOS, MAS UTILIZA OUTRA APLICAÇÃO, NÃO APRESENTA TESTES EM INSTANCIAS NA NUVEM E NÃO FAZ TESTES COMPARATIVOS
\cite{at2020italo} proposed using PATSMA with CSA to optimize the chunk size of parallel loops of a reverse time migration (RTM). Here, we expand \cite{at2020italo}'s approach to make it suitable for an FWI. FWI poses an extra challenge to auto-tuning methods as it has higher algorithmic complexity and a potentially different input for each iteration. This paper also increases reproducibility as tests were also performed in cloud instances. Finally, we present comparative tests with a state-of-the-art auto-tuning method.

\section{Conclusion} \label{section:conclusoes}

This work proposed an auto-tuning strategy based on the Parameter Auto-Tuning for Shared Memory Algorithms (PATSMA) to find an ideal chunk size for a \textit{dynamic} OpenMP scheduler. This approach was designed to work in different computing environments, regardless of parameters such as number of threads, processors, and amount of memory. We applied the proposed mechanism to a Full Waveform Inversion (FWI) algorithm, reducing its execution time.

Experiments varying the input size and in different computing environments showed that the proposed auto-tuning outperforms the default OpenMP \textit{static} and \textit{guided} schedulers in all tested scenarios. For those experiments, the proposed strategy achieved speedups of up to $70.46\%$. Those results demonstrate that our method adapts well to different problem sizes and computing environments, outperforming the OpenMP standard schedulers.

Another set of experiments applied to an FWI showed the performance of the proposed method compared to the OpenMP \textit{static} and \textit{guided} schedulers for different numbers of seismic shots. In this set of tests, the proposed auto-tuning also outperformed the OpenMP schedulers in tested scenarios, reaching up to $14.96\%$ speedup. In this set of tests, it was observed that the speedup remained stable because the overhead did not increase with the increase in the number of shots. It is possible to conclude that the chunk size adjusted for the first shot can be reused for the following shots and iterations of FWI. In all cases, the overhead was below $1.2\%$. The results also show that the proposed method is scalable, presenting higher speedups for tests with larger input sizes and on architectures with more cores.

Finally, we have compared the execution times of an FWI using our proposed auto-tuning, the default OpenMP \textit{static} and \textit{guided} schedulers, and the OpenMP \textit{static}, \textit{dynamic} and \textit{guided} schedulers with the chunk size determined by the formula proposed by \cite{auto4omp2022}. Those tests were performed in two machines for three input sizes. Our method outperformed the schedulers tested in all scenarios, reaching speedups from $2.07\%$ to $34.95\%$.

\section*{Computer code availability}

Code essential to the analysis in this paper is available on GitLab at the following public repositories: (1) \url{https://gitlab.com/lappsufrn/seismic/ufrn-fwi/mamute} Full Waveform Inversion code in a repository called Mamute and (2) \url{https://gitlab.com/lappsufrn/patsma} PATSMA code. All codes utilized in this paper were written in C++ language and are open source under the MIT License.

\section{Acknowledgments}

The authors gratefully acknowledge support from Shell Brazil through the project ``\textit{Novas metodologias computacionalmente escal\'{a}veis para s\'{i}smica 4D orientado ao alvo em reservat\'{o}rios do pr\'{e}-sal}'' at the Universidade Federal do Rio Grande do Norte (UFRN) and the strategic importance of the support given by ANP through the R\&D levy regulation.
This work was partly supported by Oracle Cloud credits and related resources the Oracle for Research program provided.
The authors also acknowledge the National Laboratory for Scientific Computing (LNCC/MCTI, Brazil) and the High-Performance Computing Center at UFRN (NPAD/UFRN) for providing HPC resources of the SDumont and NPAD supercomputers, which have contributed to the research results reported within this paper. Felipe Silva has been an undergraduate researcher through the PIVIC and PIVIC-Af programs at UFERSA.

\newpage

% \textbf{Code availability section}

% Name of the code/library

% Contact: e-mail and phone number

% Hardware requirements: ...

% Program language: ...
 
% Software required: ...

% Program size: ...

% The source codes are available for downloading at the link:
% https://github.com/ . . . . 

\bibliographystyle{cas-model2-names}
\bibliography{bibliography} 

\begin{thebibliography}{42}
\expandafter\ifx\csname natexlab\endcsname\relax\def\natexlab#1{#1}\fi
\providecommand{\url}[1]{\texttt{#1}}
\providecommand{\href}[2]{#2}
\providecommand{\path}[1]{#1}
\providecommand{\DOIprefix}{doi:}
\providecommand{\ArXivprefix}{arXiv:}
\providecommand{\URLprefix}{URL: }
\providecommand{\Pubmedprefix}{pmid:}
\providecommand{\doi}[1]{\href{http://dx.doi.org/#1}{\path{#1}}}
\providecommand{\Pubmed}[1]{\href{pmid:#1}{\path{#1}}}
\providecommand{\bibinfo}[2]{#2}
\ifx\xfnm\relax \def\xfnm[#1]{\unskip,\space#1}\fi
%Type = Incollection
\bibitem[{Andreolli et~al.(2015)Andreolli, Thierry, Borges, Skinner and
  Yount}]{Andreolli2015}
\bibinfo{author}{Andreolli, C.}, \bibinfo{author}{Thierry, P.},
  \bibinfo{author}{Borges, L.}, \bibinfo{author}{Skinner, G.},
  \bibinfo{author}{Yount, C.}, \bibinfo{year}{2015}.
\newblock \bibinfo{title}{{Characterization and Optimization Methodology
  Applied to Stencil Computations}}, in: \bibinfo{editor}{Jeffers, J.},
  \bibinfo{editor}{Reinders, J.} (Eds.), \bibinfo{booktitle}{High Performance
  Parallelism Pearls}. \bibinfo{publisher}{Elsevier}.
  chapter~\bibinfo{chapter}{23}, pp. \bibinfo{pages}{377--396}.
\newblock \URLprefix
  \url{http://www.sciencedirect.com/science/article/pii/B9780128021187000236},
  \DOIprefix\doi{10.1016/B978-0-12-802118-7.00023-6}.
%Type = Inproceedings
\bibitem[{Andreolli et~al.(2014)Andreolli, Thierry, Borges, Yount and
  Skinner}]{Andreolli2014}
\bibinfo{author}{Andreolli, C.}, \bibinfo{author}{Thierry, P.},
  \bibinfo{author}{Borges, L.}, \bibinfo{author}{Yount, C.},
  \bibinfo{author}{Skinner, G.}, \bibinfo{year}{2014}.
\newblock \bibinfo{title}{{Genetic Algorithm Based Auto-Tuning of Seismic
  Applications on Multi and Manycore Computers}}, in: \bibinfo{booktitle}{EAGE
  Workshop on High Performance Computing for Upstream}, p.~\bibinfo{pages}{0}.
\newblock \DOIprefix\doi{10.3997/2214-4609.20141920}.
%Type = Article
\bibitem[{Assis et~al.(2020)Assis, Fernandes, Barros and
  Xavier-De-Souza}]{at2020italo}
\bibinfo{author}{Assis, I.A.S.}, \bibinfo{author}{Fernandes, J.B.},
  \bibinfo{author}{Barros, T.}, \bibinfo{author}{Xavier-De-Souza, S.},
  \bibinfo{year}{2020}.
\newblock \bibinfo{title}{Auto-tuning of dynamic scheduling applied to 3d
  reverse time migration on multicore systems}.
\newblock \bibinfo{journal}{IEEE Access} \bibinfo{volume}{8},
  \bibinfo{pages}{145115--145127}.
\newblock \DOIprefix\doi{10.1109/ACCESS.2020.3015045}.
%Type = Inproceedings
\bibitem[{Bak et~al.(2018)Bak, Menon, White, Diener and Kale}]{at2018bak}
\bibinfo{author}{Bak, S.}, \bibinfo{author}{Menon, H.}, \bibinfo{author}{White,
  S.}, \bibinfo{author}{Diener, M.}, \bibinfo{author}{Kale, L.},
  \bibinfo{year}{2018}.
\newblock \bibinfo{title}{Multi-level load balancing with an integrated runtime
  approach}, in: \bibinfo{booktitle}{Proceedings of the 18th IEEE/ACM
  International Symposium on Cluster, Cloud and Grid Computing},
  \bibinfo{publisher}{IEEE Press}. p. \bibinfo{pages}{31–40}.
\newblock \URLprefix \url{https://doi.org/10.1109/CCGRID.2018.00018},
  \DOIprefix\doi{10.1109/CCGRID.2018.00018}.
%Type = Inproceedings
\bibitem[{Barros et~al.(2018)Barros, Fernandes, de~Assis and
  de~Souza}]{Barros2018}
\bibinfo{author}{Barros, T.}, \bibinfo{author}{Fernandes, J.B.},
  \bibinfo{author}{de~Assis, I.A.S.}, \bibinfo{author}{de~Souza, S.X.},
  \bibinfo{year}{2018}.
\newblock \bibinfo{title}{Auto-tuning of 3d acoustic wave propagation in shared
  memory environments}, in: \bibinfo{booktitle}{""},
  \bibinfo{publisher}{EarthDoc}. p.~\bibinfo{pages}{0}.
\newblock \URLprefix
  \url{http://www.earthdoc.org/publication/publicationdetails/?publication=94579},
  \DOIprefix\doi{10.3997/2214-4609.201803072}.
%Type = Article
\bibitem[{Baysal et~al.(1983)Baysal, Kosloff and Sherwood}]{Baysal1983}
\bibinfo{author}{Baysal, E.}, \bibinfo{author}{Kosloff, D.D.},
  \bibinfo{author}{Sherwood, J.W.C.}, \bibinfo{year}{1983}.
\newblock \bibinfo{title}{Reverse time migration}.
\newblock \bibinfo{journal}{Geophysics} \bibinfo{volume}{48},
  \bibinfo{pages}{1514--1524}.
\newblock \DOIprefix\doi{10.1190/1.1441434}.
%Type = Inproceedings
\bibitem[{Bağbaba et~al.(2021)Bağbaba, Wang, Niethammer and
  Gracia}]{at2021bagbaba}
\bibinfo{author}{Bağbaba, A.}, \bibinfo{author}{Wang, X.},
  \bibinfo{author}{Niethammer, C.}, \bibinfo{author}{Gracia, J.},
  \bibinfo{year}{2021}.
\newblock \bibinfo{title}{Improving the i/o performance of applications with
  predictive modeling based auto-tuning}, in: \bibinfo{booktitle}{2021
  International Conference on Engineering and Emerging Technologies (ICEET)},
  pp. \bibinfo{pages}{1--6}.
\newblock \DOIprefix\doi{10.1109/ICEET53442.2021.9659711}.
%Type = Phdthesis
\bibitem[{Bez(2021)}]{at2021bez}
\bibinfo{author}{Bez, J.L.}, \bibinfo{year}{2021}.
\newblock \bibinfo{title}{Dynamic Tuning and Reconfiguration of the I/O
  Forwarding Layer in HPC Platforms}.
\newblock Ph.D. thesis. 'Universidade Federal do Rio Grande do Sul'.
\newblock \DOIprefix\doi{10.13140/RG.2.2.18591.48802}.
%Type = Misc
\bibitem[{Board(2021)}]{OpenMP-spec}
\bibinfo{author}{Board, O.A.R.}, \bibinfo{year}{2021}.
\newblock \bibinfo{title}{{OpenMP} specifications}.
\newblock \URLprefix \url{https://www.openmp.org/specifications/}.
  \bibinfo{note}{version 5.1}.
%Type = Article
\bibitem[{{Dagum} and {Menon}(1998)}]{OpemMP}
\bibinfo{author}{{Dagum}, L.}, \bibinfo{author}{{Menon}, R.},
  \bibinfo{year}{1998}.
\newblock \bibinfo{title}{Open{MP}: an industry standard {API} for
  shared-memory programming}.
\newblock \bibinfo{journal}{IEEE Computational Science and Engineering}
  \bibinfo{volume}{5}, \bibinfo{pages}{46--55}.
\newblock \DOIprefix\doi{10.1109/99.660313}.
%Type = Article
\bibitem[{Dennis~Booth and Allen~Lane(2022)}]{ich2022booth}
\bibinfo{author}{Dennis~Booth, J.}, \bibinfo{author}{Allen~Lane, P.},
  \bibinfo{year}{2022}.
\newblock \bibinfo{title}{An adaptive self-scheduling loop scheduler}.
\newblock \bibinfo{journal}{Concurrency and Computation: Practice and
  Experience} \bibinfo{volume}{34}, \bibinfo{pages}{e6750}.
\newblock \URLprefix
  \url{https://onlinelibrary.wiley.com/doi/abs/10.1002/cpe.6750},
  \DOIprefix\doi{https://doi.org/10.1002/cpe.6750},
  \href{http://arxiv.org/abs/https://onlinelibrary.wiley.com/doi/pdf/10.1002/cpe.6750}{\tt
  arXiv:https://onlinelibrary.wiley.com/doi/pdf/10.1002/cpe.6750}.
%Type = Inproceedings
\bibitem[{Dutta et~al.(2022)Dutta, Alcaraz, TehraniJamsaz, Sikora, Cesar and
  Jannesari}]{at2022dutta}
\bibinfo{author}{Dutta, A.}, \bibinfo{author}{Alcaraz, J.},
  \bibinfo{author}{TehraniJamsaz, A.}, \bibinfo{author}{Sikora, A.},
  \bibinfo{author}{Cesar, E.}, \bibinfo{author}{Jannesari, A.},
  \bibinfo{year}{2022}.
\newblock \bibinfo{title}{Pattern-based autotuning of openmp loops using graph
  neural networks}, in: \bibinfo{booktitle}{2022 IEEE/ACM International
  Workshop on Artificial Intelligence and Machine Learning for Scientific
  Applications (AI4S)}, pp. \bibinfo{pages}{26--31}.
\newblock \DOIprefix\doi{10.1109/AI4S56813.2022.00010}.
%Type = Inproceedings
\bibitem[{Fernandes et~al.(2018)Fernandes, \'{I}talo A. Souza-de Assis, Barros
  and de~Souza}]{Fernandes2018}
\bibinfo{author}{Fernandes, J.B.}, \bibinfo{author}{\'{I}talo A. Souza-de
  Assis}, \bibinfo{author}{Barros, T.}, \bibinfo{author}{de~Souza, S.X.},
  \bibinfo{year}{2018}.
\newblock \bibinfo{title}{Automatic scheduler for 3d seismic modeling by finite
  differences}, in: \bibinfo{booktitle}{''}, p.~\bibinfo{pages}{0}.
\newblock \URLprefix
  \url{https://stt.ibp.org.br/eventos/2018/riooil2018/pdfs/Riooil2018_1901_201806151345riooeg_end_paper.pdf}.
%Type = Misc
\bibitem[{Fernandes et~al.(2025)Fernandes, Oliveira, Silva, da~Silva,
  Rodrigues, Schneider, Bianchini, de~Araujo, Barros, Ítalo A.~S.~Assis and
  de~Souza}]{mamute2025preprint}
\bibinfo{author}{Fernandes, J.B.}, \bibinfo{author}{Oliveira, A.D.S.},
  \bibinfo{author}{Silva, M.C.A.T.}, \bibinfo{author}{da~Silva, F.H.S.},
  \bibinfo{author}{Rodrigues, V.H.M.}, \bibinfo{author}{Schneider, K.A.},
  \bibinfo{author}{Bianchini, C.P.}, \bibinfo{author}{de~Araujo, J.M.},
  \bibinfo{author}{Barros, T.}, \bibinfo{author}{Ítalo A.~S.~Assis},
  \bibinfo{author}{de~Souza, S.X.}, \bibinfo{year}{2025}.
\newblock \bibinfo{title}{Mamute: high-performance computing for geophysical
  methods}.
\newblock \URLprefix \url{https://arxiv.org/abs/2502.12350},
  \href{http://arxiv.org/abs/2502.12350}{\tt arXiv:2502.12350}.
%Type = Misc
\bibitem[{Fernandes et~al.(2024)Fernandes, da~Silva, de~Souza and
  Assis}]{fernandes2024patsma}
\bibinfo{author}{Fernandes, J.B.}, \bibinfo{author}{da~Silva, F.H.S.},
  \bibinfo{author}{de~Souza, S.X.}, \bibinfo{author}{Assis, I.A.S.},
  \bibinfo{year}{2024}.
\newblock \bibinfo{title}{Patsma: Parameter auto-tuning for shared memory
  algorithms}.
\newblock \href{http://arxiv.org/abs/2401.07861}{\tt arXiv:2401.07861}.
%Type = Article
\bibitem[{Griewank and Walther(2000)}]{checkpoint2000griewank}
\bibinfo{author}{Griewank, A.}, \bibinfo{author}{Walther, A.},
  \bibinfo{year}{2000}.
\newblock \bibinfo{title}{Algorithm 799: Revolve: An implementation of
  checkpointing for the reverse or adjoint mode of computational
  differentiation}.
\newblock \bibinfo{journal}{ACM Trans. Math. Softw.} \bibinfo{volume}{26},
  \bibinfo{pages}{19–45}.
\newblock \URLprefix \url{https://doi.org/10.1145/347837.347846},
  \DOIprefix\doi{10.1145/347837.347846}.
%Type = Inproceedings
\bibitem[{HoseinyFarahabady et~al.(2020)HoseinyFarahabady, Taheri, Zomaya and
  Tari}]{at2020hose}
\bibinfo{author}{HoseinyFarahabady, M.R.}, \bibinfo{author}{Taheri, J.},
  \bibinfo{author}{Zomaya, A.Y.}, \bibinfo{author}{Tari, Z.},
  \bibinfo{year}{2020}.
\newblock \bibinfo{title}{Auto-tuning of large-scale iterative operations on
  modern streaming platforms}, in: \bibinfo{booktitle}{Proceedings of the 16th
  International Conference on Emerging Networking EXperiments and
  Technologies}, \bibinfo{publisher}{Association for Computing Machinery},
  \bibinfo{address}{New York, NY, USA}. p. \bibinfo{pages}{554–555}.
\newblock \URLprefix \url{https://doi.org/10.1145/3386367.3431680},
  \DOIprefix\doi{10.1145/3386367.3431680}.
%Type = Inproceedings
\bibitem[{Kale et~al.(2014)Kale, Randles and Gropp}]{Kale2014}
\bibinfo{author}{Kale, V.}, \bibinfo{author}{Randles, A.},
  \bibinfo{author}{Gropp, W.D.}, \bibinfo{year}{2014}.
\newblock \bibinfo{title}{Locality-optimized mixed static/dynamic scheduling
  for improving load balancing on {SMP}s}, in: \bibinfo{booktitle}{Proceedings
  of the 21st European MPI Users' Group Meeting}, pp.
  \bibinfo{pages}{115--116}.
%Type = Inproceedings
\bibitem[{Katagiri et~al.(2014)Katagiri, Ohshima and Matsumoto}]{Katagiri2014}
\bibinfo{author}{Katagiri, T.}, \bibinfo{author}{Ohshima, S.},
  \bibinfo{author}{Matsumoto, M.}, \bibinfo{year}{2014}.
\newblock \bibinfo{title}{{Auto-tuning of computation kernels from an FDM code
  with ppOpen-AT}}, in: \bibinfo{booktitle}{Proceedings - 2014 IEEE 8th
  International Symposium on Embedded Multicore/Manycore SoCs, MCSoC 2014},
  p.~\bibinfo{pages}{0}.
\newblock \DOIprefix\doi{10.1109/MCSoC.2014.22}.
%Type = Inproceedings
\bibitem[{Katagiri et~al.(2015)Katagiri, Ohshima and Matsumoto}]{Katagiri2015}
\bibinfo{author}{Katagiri, T.}, \bibinfo{author}{Ohshima, S.},
  \bibinfo{author}{Matsumoto, M.}, \bibinfo{year}{2015}.
\newblock \bibinfo{title}{{Directive-Based Auto-Tuning for the Finite
  Difference Method on the Xeon Phi}}, in: \bibinfo{booktitle}{Proceedings -
  2015 IEEE 29th International Parallel and Distributed Processing Symposium
  Workshops, IPDPSW 2015}, p.~\bibinfo{pages}{0}.
\newblock \DOIprefix\doi{10.1109/IPDPSW.2015.11}.
%Type = Article
\bibitem[{Kimovski et~al.(2021)Kimovski, Math{\'a}, Iuhasz, Marozzo, Petcu and
  Prodan}]{kimovski2021autotuning}
\bibinfo{author}{Kimovski, D.}, \bibinfo{author}{Math{\'a}, R.},
  \bibinfo{author}{Iuhasz, G.}, \bibinfo{author}{Marozzo, F.},
  \bibinfo{author}{Petcu, D.}, \bibinfo{author}{Prodan, R.},
  \bibinfo{year}{2021}.
\newblock \bibinfo{title}{Autotuning of exascale applications with anomalies
  detection}.
\newblock \bibinfo{journal}{Frontiers in Big Data} \bibinfo{volume}{4},
  \bibinfo{pages}{657218}.
%Type = Article
\bibitem[{Kirkpatrick et~al.(1983)Kirkpatrick, Gelatt and
  Vecchi}]{sa1983kirkpatrick}
\bibinfo{author}{Kirkpatrick, S.}, \bibinfo{author}{Gelatt, C.D.},
  \bibinfo{author}{Vecchi, M.P.}, \bibinfo{year}{1983}.
\newblock \bibinfo{title}{Optimization by simulated annealing}.
\newblock \bibinfo{journal}{Science} \bibinfo{volume}{220},
  \bibinfo{pages}{671--680}.
\newblock \DOIprefix\doi{10.1126/science.220.4598.671}.
%Type = Article
\bibitem[{Kruse et~al.(2020)Kruse, Finkel and Wu}]{at2020kruse}
\bibinfo{author}{Kruse, M.}, \bibinfo{author}{Finkel, H.}, \bibinfo{author}{Wu,
  X.}, \bibinfo{year}{2020}.
\newblock \bibinfo{title}{Autotuning search space for loop transformations}.
\newblock \bibinfo{journal}{CoRR} \bibinfo{volume}{abs/2010.06521}.
\newblock \URLprefix \url{https://arxiv.org/abs/2010.06521},
  \href{http://arxiv.org/abs/2010.06521}{\tt arXiv:2010.06521}.
%Type = Inproceedings
\bibitem[{Liu et~al.(2021)Liu, Sid-Lakhdar, Marques, Zhu, Meng, Demmel and
  Li}]{Gptune2021liu}
\bibinfo{author}{Liu, Y.}, \bibinfo{author}{Sid-Lakhdar, W.M.},
  \bibinfo{author}{Marques, O.}, \bibinfo{author}{Zhu, X.},
  \bibinfo{author}{Meng, C.}, \bibinfo{author}{Demmel, J.W.},
  \bibinfo{author}{Li, X.S.}, \bibinfo{year}{2021}.
\newblock \bibinfo{title}{Gptune: Multitask learning for autotuning exascale
  applications}, in: \bibinfo{booktitle}{GPTune: Multitask Learning for
  Autotuning Exascale Applications}, \bibinfo{publisher}{Association for
  Computing Machinery}, \bibinfo{address}{New York, NY, USA}. p.
  \bibinfo{pages}{234–246}.
\newblock \URLprefix \url{https://doi.org/10.1145/3437801.3441621},
  \DOIprefix\doi{10.1145/3437801.3441621}.
%Type = Inproceedings
\bibitem[{Menon et~al.(2020)Menon, Bhatele and Gamblin}]{hiperbot2020menon}
\bibinfo{author}{Menon, H.}, \bibinfo{author}{Bhatele, A.},
  \bibinfo{author}{Gamblin, T.}, \bibinfo{year}{2020}.
\newblock \bibinfo{title}{Auto-tuning parameter choices in hpc applications
  using bayesian optimization}, in: \bibinfo{booktitle}{2020 IEEE International
  Parallel and Distributed Processing Symposium (IPDPS)}, pp.
  \bibinfo{pages}{831--840}.
\newblock \DOIprefix\doi{10.1109/IPDPS47924.2020.00090}.
%Type = Phdthesis
\bibitem[{Milani(2020)}]{at2020milani}
\bibinfo{author}{Milani, L.}, \bibinfo{year}{2020}.
\newblock \bibinfo{title}{Autotuning with machine learning of OpenMP task
  applications}.
\newblock Ph.D. thesis. 'Université Grenoble Alpes'.
\newblock \URLprefix \url{https://theses.hal.science/tel-03227414}.
%Type = Article
\bibitem[{Mohammed et~al.(2022)Mohammed, Korndörfer, Eleliemy and
  Ciorba}]{auto4omp2022}
\bibinfo{author}{Mohammed, A.}, \bibinfo{author}{Korndörfer, J.H.M.},
  \bibinfo{author}{Eleliemy, A.}, \bibinfo{author}{Ciorba, F.M.},
  \bibinfo{year}{2022}.
\newblock \bibinfo{title}{Automated scheduling algorithm selection and chunk
  parameter calculation in openmp}.
\newblock \bibinfo{journal}{IEEE Transactions on Parallel and Distributed
  Systems} \bibinfo{volume}{33}, \bibinfo{pages}{4383--4394}.
\newblock \DOIprefix\doi{10.1109/TPDS.2022.3189270}.
%Type = Inproceedings
\bibitem[{Padoin et~al.(2014)Padoin, Castro, Pilla, Navaux and
  M{é}haut}]{Padoin2014}
\bibinfo{author}{Padoin, E.L.}, \bibinfo{author}{Castro, M.},
  \bibinfo{author}{Pilla, L.L.}, \bibinfo{author}{Navaux, P.O.A.},
  \bibinfo{author}{M{é}haut, J.}, \bibinfo{year}{2014}.
\newblock \bibinfo{title}{{Saving energy by exploiting residual imbalances on
  iterative applications}}, in: \bibinfo{booktitle}{2014 21st International
  Conference on High Performance Computing (HiPC)}, pp. \bibinfo{pages}{1--10}.
\newblock \DOIprefix\doi{10.1109/HiPC.2014.7116895}.
%Type = Inproceedings
\bibitem[{Padoin et~al.(2017)Padoin, Pilla, Castro, Navaux and
  M{é}haut}]{Padoin2017}
\bibinfo{author}{Padoin, E.L.}, \bibinfo{author}{Pilla, L.L.},
  \bibinfo{author}{Castro, M.}, \bibinfo{author}{Navaux, P.O.},
  \bibinfo{author}{M{é}haut, J.F.}, \bibinfo{year}{2017}.
\newblock \bibinfo{title}{{Exploration of load balancing thresholds to save
  energy on iterative applications}}, in: \bibinfo{booktitle}{Communications in
  Computer and Information Science}, p.~\bibinfo{pages}{0}.
\newblock \DOIprefix\doi{10.1007/978-3-319-57972-6_6}.
%Type = Article
\bibitem[{Plessix(2006)}]{plessix2006adjoint}
\bibinfo{author}{Plessix, R.E.}, \bibinfo{year}{2006}.
\newblock \bibinfo{title}{A review of the adjoint-state method for computing
  the gradient of a functional with geophysical applications}.
\newblock \bibinfo{journal}{Geophysical Journal International}
  \bibinfo{volume}{167}, \bibinfo{pages}{495--503}.
\newblock \URLprefix
  \url{https://onlinelibrary.wiley.com/doi/abs/10.1111/j.1365-246X.2006.02978.x},
  \DOIprefix\doi{https://doi.org/10.1111/j.1365-246X.2006.02978.x},
  \href{http://arxiv.org/abs/https://onlinelibrary.wiley.com/doi/pdf/10.1111/j.1365-246X.2006.02978.x}{\tt
  arXiv:https://onlinelibrary.wiley.com/doi/pdf/10.1111/j.1365-246X.2006.02978.x}.
%Type = Article
\bibitem[{Rasch et~al.(2021)Rasch, Schulze, Steuwer and Gorlatch}]{at2021rash}
\bibinfo{author}{Rasch, A.}, \bibinfo{author}{Schulze, R.},
  \bibinfo{author}{Steuwer, M.}, \bibinfo{author}{Gorlatch, S.},
  \bibinfo{year}{2021}.
\newblock \bibinfo{title}{Efficient auto-tuning of parallel programs with
  interdependent tuning parameters via auto-tuning framework (atf)}.
\newblock \bibinfo{journal}{Association for Computing Machinery}
  \bibinfo{volume}{18}.
\newblock \URLprefix \url{https://doi.org/10.1145/3427093},
  \DOIprefix\doi{10.1145/3427093}.
%Type = Inproceedings
\bibitem[{Robert et~al.(2020)Robert, Zertal and Couvee}]{shaman2021robert}
\bibinfo{author}{Robert, S.}, \bibinfo{author}{Zertal, S.},
  \bibinfo{author}{Couvee, P.}, \bibinfo{year}{2020}.
\newblock \bibinfo{title}{Shaman: A flexible framework for auto-tuning hpc
  systems}, in: \bibinfo{editor}{Calzarossa, M.C.}, \bibinfo{editor}{Gelenbe,
  E.}, \bibinfo{editor}{Grochla, K.}, \bibinfo{editor}{Lent, R.},
  \bibinfo{editor}{Czachórski, T.} (Eds.), \bibinfo{booktitle}{Modelling,
  Analysis, and Simulation of Computer and Telecommunication Systems - 28th
  International Symposium, MASCOTS 2020, Nice, France, November 17-19, 2020,
  Revised Selected Papers}, \bibinfo{publisher}{Springer}. pp.
  \bibinfo{pages}{147--158}.
\newblock \URLprefix \url{https://doi.org/10.1007/978-3-030-68110-4_10},
  \DOIprefix\doi{10.1007/978-3-030-68110-4_10}.
%Type = Inproceedings
\bibitem[{Rocha et~al.(2020)Rocha, Schwarzrock, Pereira, Schnorr, Navaux,
  Lorenzon and Filho}]{at2020Hiago}
\bibinfo{author}{Rocha, H.}, \bibinfo{author}{Schwarzrock, J.},
  \bibinfo{author}{Pereira, M.}, \bibinfo{author}{Schnorr, L.},
  \bibinfo{author}{Navaux, P.}, \bibinfo{author}{Lorenzon, A.},
  \bibinfo{author}{Filho, A.C.B.}, \bibinfo{year}{2020}.
\newblock \bibinfo{title}{Attune: A heuristic based framework for parallel
  applications autotuning}, in: \bibinfo{booktitle}{Anais Estendidos do X
  Simpósio Brasileiro de Engenharia de Sistemas Computacionais},
  \bibinfo{publisher}{SBC}, \bibinfo{address}{Porto Alegre, RS, Brasil}. pp.
  \bibinfo{pages}{151--156}.
\newblock \URLprefix
  \url{https://sol.sbc.org.br/index.php/sbesc_estendido/article/view/13105},
  \DOIprefix\doi{10.5753/sbesc_estendido.2020.13105}.
%Type = Inproceedings
\bibitem[{Roy et~al.(2021)Roy, Patel, Gadepally and Tiwari}]{at2021roy}
\bibinfo{author}{Roy, R.B.}, \bibinfo{author}{Patel, T.},
  \bibinfo{author}{Gadepally, V.}, \bibinfo{author}{Tiwari, D.},
  \bibinfo{year}{2021}.
\newblock \bibinfo{title}{Bliss: Auto-tuning complex applications using a pool
  of diverse lightweight learning models}, in: \bibinfo{booktitle}{Proceedings
  of the 42nd ACM SIGPLAN International Conference on Programming Language
  Design and Implementation}, \bibinfo{publisher}{Association for Computing
  Machinery}, \bibinfo{address}{New York, NY, USA}. p.
  \bibinfo{pages}{1280–1295}.
\newblock \URLprefix \url{https://doi.org/10.1145/3453483.3454109},
  \DOIprefix\doi{10.1145/3453483.3454109}.
%Type = Article
\bibitem[{Sakurai et~al.(2020)Sakurai, Katagiri, Ohshima and
  Nagai}]{at2020sakurai}
\bibinfo{author}{Sakurai, T.}, \bibinfo{author}{Katagiri, T.},
  \bibinfo{author}{Ohshima, S.}, \bibinfo{author}{Nagai, T.},
  \bibinfo{year}{2020}.
\newblock \bibinfo{title}{Autotuning by changing directives and number of
  threads in openmp using ppopen-at}.
\newblock \bibinfo{journal}{''} \URLprefix
  \url{http://rgdoi.net/10.13140/RG.2.2.26988.80005},
  \DOIprefix\doi{10.13140/RG.2.2.26988.80005}.
%Type = Inproceedings
\bibitem[{Seiferth et~al.(2020)Seiferth, Korch and Rauber}]{at2020seiferth}
\bibinfo{author}{Seiferth, J.}, \bibinfo{author}{Korch, M.},
  \bibinfo{author}{Rauber, T.}, \bibinfo{year}{2020}.
\newblock \bibinfo{title}{Offsite autotuning approach: Performance model driven
  autotuning applied to parallel explicit ode methods}, in:
  \bibinfo{booktitle}{High Performance Computing: 35th International
  Conference, ISC High Performance 2020, Frankfurt/Main, Germany, June 22–25,
  2020, Proceedings}, \bibinfo{publisher}{Springer-Verlag},
  \bibinfo{address}{Berlin, Heidelberg}. p. \bibinfo{pages}{370–390}.
\newblock \URLprefix \url{https://doi.org/10.1007/978-3-030-50743-5_19},
  \DOIprefix\doi{10.1007/978-3-030-50743-5_19}.
%Type = Inproceedings
\bibitem[{Shu et~al.(2021)Shu, Guo, Wozniak, Ding, Foster and Kurc}]{Shu2021}
\bibinfo{author}{Shu, T.}, \bibinfo{author}{Guo, Y.}, \bibinfo{author}{Wozniak,
  J.}, \bibinfo{author}{Ding, X.}, \bibinfo{author}{Foster, I.},
  \bibinfo{author}{Kurc, T.}, \bibinfo{year}{2021}.
\newblock \bibinfo{title}{Bootstrapping in-situ workflow auto-tuning via
  combining performance models of component applications}, in:
  \bibinfo{booktitle}{Proceedings of the International Conference for High
  Performance Computing, Networking, Storage and Analysis},
  \bibinfo{publisher}{ACM}. pp. \bibinfo{pages}{1--15}.
\newblock \URLprefix \url{https://dl.acm.org/doi/10.1145/3458817.3476197},
  \DOIprefix\doi{10.1145/3458817.3476197}.
%Type = Article
\bibitem[{Gon{\c{c}}alves-e Silva et~al.(2018)Gon{\c{c}}alves-e Silva, Aloise
  and Xavier-de Souza}]{gonccalves2018parallel}
\bibinfo{author}{Gon{\c{c}}alves-e Silva, K.}, \bibinfo{author}{Aloise, D.},
  \bibinfo{author}{Xavier-de Souza, S.}, \bibinfo{year}{2018}.
\newblock \bibinfo{title}{Parallel synchronous and asynchronous coupled
  simulated annealing}.
\newblock \bibinfo{journal}{The Journal of Supercomputing}
  \bibinfo{volume}{74}, \bibinfo{pages}{2841--2869}.
%Type = Article
\bibitem[{Xavier-de Souza et~al.(2010)Xavier-de Souza, Suykens, Vandewalle and
  Bolle}]{csa2010samuel}
\bibinfo{author}{Xavier-de Souza, S.}, \bibinfo{author}{Suykens, J.A.K.},
  \bibinfo{author}{Vandewalle, J.}, \bibinfo{author}{Bolle, D.},
  \bibinfo{year}{2010}.
\newblock \bibinfo{title}{Coupled simulated annealing}.
\newblock \bibinfo{journal}{IEEE Transactions on Systems, Man, and Cybernetics,
  Part B (Cybernetics)} \bibinfo{volume}{40}, \bibinfo{pages}{320--335}.
\newblock \DOIprefix\doi{10.1109/TSMCB.2009.2020435}.
%Type = Article
\bibitem[{Symes(2007)}]{symes2010reverse}
\bibinfo{author}{Symes, W.W.}, \bibinfo{year}{2007}.
\newblock \bibinfo{title}{Reverse time migration with optimal checkpointing}.
\newblock \bibinfo{journal}{GEOPHYSICS} \bibinfo{volume}{72},
  \bibinfo{pages}{SM213--SM221}.
\newblock \URLprefix \url{https://doi.org/10.1190/1.2742686},
  \DOIprefix\doi{10.1190/1.2742686},
  \href{http://arxiv.org/abs/https://doi.org/10.1190/1.2742686}{\tt
  arXiv:https://doi.org/10.1190/1.2742686}.
%Type = Article
\bibitem[{Tarantola(1984)}]{tarantola1984inversion}
\bibinfo{author}{Tarantola, A.}, \bibinfo{year}{1984}.
\newblock \bibinfo{title}{Inversion of seismic reflection data in the acoustic
  approximation}.
\newblock \bibinfo{journal}{Geophysics} \bibinfo{volume}{49},
  \bibinfo{pages}{1259--1266}.
\newblock \DOIprefix\doi{10.1190/1.1441744}.
%Type = Inproceedings
\bibitem[{Wood et~al.(2021)Wood, Georgakoudis, Beckingsale, Poliakoff, Gimenez,
  Huck, Malony and Gamblin}]{at2021wood}
\bibinfo{author}{Wood, C.}, \bibinfo{author}{Georgakoudis, G.},
  \bibinfo{author}{Beckingsale, D.}, \bibinfo{author}{Poliakoff, D.},
  \bibinfo{author}{Gimenez, A.}, \bibinfo{author}{Huck, K.},
  \bibinfo{author}{Malony, A.}, \bibinfo{author}{Gamblin, T.},
  \bibinfo{year}{2021}.
\newblock \bibinfo{title}{Artemis: Automatic runtime tuning of parallel
  execution parameters using machine learning}, in: \bibinfo{booktitle}{High
  Performance Computing: 36th International Conference, ISC High Performance
  2021, Virtual Event, June 24 – July 2, 2021, Proceedings},
  \bibinfo{publisher}{Springer-Verlag}, \bibinfo{address}{Berlin, Heidelberg}.
  p. \bibinfo{pages}{453–472}.
\newblock \URLprefix \url{https://doi.org/10.1007/978-3-030-78713-4_24},
  \DOIprefix\doi{10.1007/978-3-030-78713-4_24}.

\end{thebibliography}

\end{document}